\documentclass[12pt]{iopart}
\pdfoutput=1
\usepackage{amsfonts}
\usepackage{graphicx}

\usepackage{xcolor}
\usepackage{subfigure}
\usepackage{textcomp}
\usepackage{cite}
\usepackage{float}
\usepackage{soul}
\usepackage{stackrel}

\definecolor{dgreen}{rgb}{0,0.7,0}

\expandafter\let\csname equation*\endcsname\relax
\expandafter\let\csname endequation*\endcsname\relax
\usepackage{amsmath,amssymb}

\pdfoutput=1
\usepackage{esint}

\usepackage{color}
\definecolor{dgreen}{rgb}{0,0.7,0}

\begin{document}
\title[]{Generalised `Arcsine' laws for run-and-tumble particle in one dimension}
\author{Prashant Singh,~Anupam Kundu}
\address{International Centre for Theoretical Sciences, TIFR, Bengaluru 560089, India}

\date{\today}

\begin{abstract}
\noindent
The 'Arcsine' laws of Brownian particles in one dimension describe distributions of three quantities: the time $t_m$ to reach maximum position, the time $t_r$ spent on the positive side and the time $t_\ell$ of the last visit to the origin. Interestingly, the cumulative distribution of all the three quantities are same and given by Arcsine function. In this paper, we study distribution of these three times $t_m,~t_r$ and $t_\ell$ in the context of single  run-and-tumble particle in one dimension, which is a simple non-Markovian process. We compute exact distributions of these three quantities for arbitrary time and find that all three distributions have delta function part and a non-delta function part. Interestingly, we find that  the distributions of $t_m$ and $t_r$ are identical (reminiscent of the Brownian particle case) when the initial velocities of the particle are chosen with equal probability. On the other hand, for $t_\ell$, only the non-delta function part is same with the other two. In addition, we find explicit expressions of the joint distributions of the maximum displacement and the time at which this maxima occurs. We verify all our analytical results through numerical simulations.
\end{abstract}

\section{Introduction}
\noindent
The `Arcsine' laws provide statistical characterisation of certain temporal behaviours of Brownian motion or more generally of discrete random processes \cite{feller68, levy39}. These laws describe the probability densities of the following three quantities measured along a trajectory over a time interval $t$ : (i) the time $t_{m}$ to reach the maximum of the process, (ii) the residence time $t_{r}$ spent on the positive (or the negative) semi axis and (iii) the last time 
$t_\ell$ the process changes sign (or crosses the origin). Remarkably, for a one dimensional Brownian process $B(\tau)$ starting from the origin, the cumulative probabilities of the above mentioned three observables are same and it is given by $F_c(\tau,t)=\text{Prob.}(t_c \leq \tau)=(2/\pi)\text{Arcsine}(\sqrt{\tau/t})$ where $t_c$ representing $t_m$, $t_r$ or $t_\ell$. These results are often known as $1^{st}$, $2^{nd}$ and $3^{rd}$ `Arcsine' laws, respectively, for obvious reason. The fact that the corresponding probability density function \begin{align}
P_{Br}(t_a,t)=\frac{1}{\pi} \frac{1}{\sqrt{t_a(t-t_a)}}, \label{P_br}
\end{align}
has counterintuitive (integrable) divergences at the edges $t_a=0$ and $t_a=t$, has  attracted a lot of general interests in these laws over the years. Since the discovery of these laws, these three quantities have been studied, either together or individually in various contexts starting from simple and constrained Brownian motion \cite{Majumdar005, Majumdar08}, queuing process \cite{Randon-Furling07}, random acceleration process, \cite{Majumdar10, Boutcheng16}, fractional Brownian motion \cite{Sadhu18}, renewal processes \cite{Baldassarri99, Godreche01, Burov11}, convex hull problems \cite{Randon-Furling09, Majumdar10con},  vicious walkers \cite{Rambeau11}, Bessel Process \cite{Schehr10} to general Markovian \cite{Dhar99, Kasahara77} and non Markovian \cite{Majumdar02, Lamperti58} processes. 
The `Arcsine' laws and related distributions have also been explored  in other  contexts like, finance \cite{Charles80, Shiryaev02, Majumdar08F}, diffusion in disordered potential \cite{SMajumdar02, Sabhapandit06}, conductance in disordered materials \cite{Nazarov94, Beenakker97}, chaotic dynamical systems \cite{Akimoto08}, quantum chaotic scattering \cite{Carlos11}, partial melting of polymers \cite{Oshanian09} and, quite recently in stochastic thermodynamics~\cite{Barato18}.

\begin{figure}[t]
\includegraphics[scale=0.4]{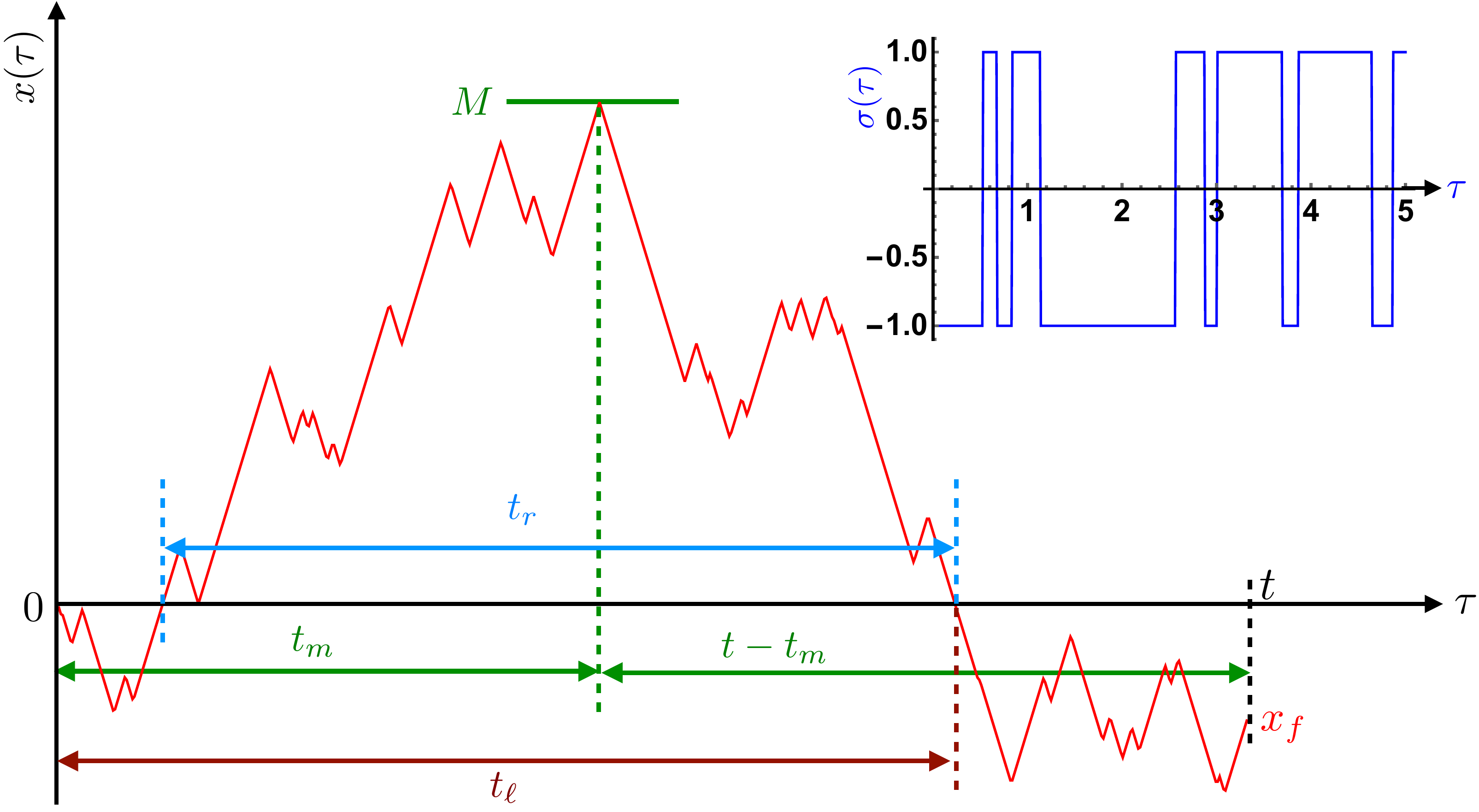}
\centering
\caption{A typical trajectory $x(\tau)$ of duration $t$ of the RTP starting from the origin. The maximum displacement $M$ occurs at time $t_m~(\leq t)$ within time interval $[0,t]$. The time duration $t_r~(\leq t)$ represents the total amount of time the RTP spends on the $x\geq 0$ region over the duration $t$. The time $t_\ell~(\leq t)$ represents the time at which the RTP crosses the origin (from below in the particular trajectory shown) for the last time and after that in the remaining time $t-t_\ell$ it did not cross the origin again. {\bf Inset:} Shows a typical realization of the telegraphic noise $\sigma(\tau)$ over interval $[0.t]$.}
\label{fig:traj}
\end{figure}

In this paper, we study the three observables, namely, $t_m$ (the time to reach maxima), $t_r$ (the residence time) and $t_\ell$ ( the time of the last visit to the origin) for a single run and tumble process. This process describes the motion of a particle in one dimension subject to telegraphic noise. The Langevin equation of motion of a run and tumble particle (RTP) is given by
\begin{align}
\frac{dx}{dt}=v\sigma (t),
\label{RTP-motion}
\end{align} 
where $x$ is the position of the particle at time $t$ and $v(>0)$ is the speed. In contrast to the Brownian motion the noise $\sigma(t)$ takes only $\pm 1$ values and changes sign with rate $\gamma$. 
The noise $\sigma(t)$ at different times are exponentially correlated $ < \sigma(t_1) \sigma(t_2)> = v^2 e^{-\gamma|t_1-t_2|}$ which makes the process $x(t)$ non-Markovian. However, in the limit $v\to \infty$ and $\gamma \to \infty$ limit keeping $\frac{v^2}{\gamma}$ finite, the noise $\eta(t)=v\sigma(t)$ becomes delta correlated $ < \eta(t_1) \eta(t_2)>= 2D \delta (t_1-t_2)$ with $D=\frac{v^2}{2 \gamma}$ and the process reduces to the usual Brownian motion. 
In Fig.~\ref{fig:traj} we show a typical trajectory $x(t)$ of a RTP particle and in the inset we show a realization of the noise $\sigma(t)$. The three observables $t_m$, $t_r$ and $t_\ell$ have also been indicated in the Fig.~\ref{fig:traj} along with their definition given in the caption. 
In the random walk literature the RTP process has been studied extensively in the past and is known as persistent random walk or telegrapher's equation \cite{Weiss02, Masoliver17}. Recently, it has received renewed interest because of its application in biological system or, more generally in active systems. 
In many theoretical studies of active matter or collective behavior of many active agents, often RTPs are used as microscopic constituents \cite{Tailleur08, Solon15}. For example E. Coli bacteria runs for some time along a straight line  and then tumbles to randomly choose a new direction of run \cite{Berg14}. In the last few years there has been  a lot of interests in studying RTPs at the individual level as they show interesting phenomena like, accumulation near boundaries \cite{Kanaya18, Tailleur09, Dhar19}, clustering \cite{Slowman16, Slowman17, Peruani12}, passive to active transition \cite{Takatori16, Kanaya19}, climbing against the hill \cite{Dauchot19}, Kramer's escape problem \cite{Thibaut18} etc.

While the single Brownian motion has been studied quite extensively over many years and a huge number of exact results related to various statistical properties of it have been proved, very less amount of study has been performed for RTPs or in general for non-Markovian processes except for a few \cite{Sadhu18, Majumdar10, Kasahara77, Majumdar02, Lamperti58}. In an attempt towards this direction, we here study the probability distribution functions $P_M(t_m,t)$, $P_R(t_r,t)$ and $P_L(t_\ell,t)$ of $t_m$, $t_r$ and $t_\ell$, respectively, measured over RTP trajectories of duration $t$ which start at the origin. We find that the distributions of all the three quantities have delta function parts which appear either at both edges or only at $t_a=0$ and, have a regular non-delta function part over $0<t_a<t$. Remarkably, we find that the non-delta function parts of the three distributions $P_M(t_m,t)$, $P_R(t_r,t)$ and $P_L(t_\ell,t)$ have identical functional form when the initial velocity directions $\pm$ are chosen with equal probability. More explicitly, we find 
\begin{align}
\begin{split}
P_M(t_m,t) &= \frac{h(t)}{2}\Big[\delta(t_m)+ \delta(t_m-t)\Big]~+~\frac{\gamma}{2}~h(t_m)h(t-t_m)  \\
P_R(t_r,t) &= \frac{h(t)}{2}\Big[\delta(t_r)+ \delta(t_r-t)\Big]~+~\frac{\gamma}{2}~h(t_r)h(t-t_r) \\
P_L(t_\ell,t) &= h(t)~\delta(t_\ell)~+~\frac{\gamma}{2}~h(t_\ell)h(t-t_\ell),~~~\text{where,} 
\end{split}
\label{results} \\
h(t) &=e^{-\gamma t} \big[I_0(\gamma t) +I_1(\gamma t) \big], \label{h(t)}     
\end{align}
and, $I_0(z)$ and $I_1(z)$ are modified Bessel functions of the first kind.
Note that the distributions $P_M(t_m,t)$ and $P_R(t_r,t)$ are identical whereas $P_L(t_\ell,t)$ is little different. It has delta function only at $t_\ell=0$. Also note that the above distributions are completely independent of the speed $v$. This is due to the fact that by definition the values of the observables $t_m$, $t_r$ and $t_\ell$ measured along a trajectory $\{x(\tau);0\leq \tau \leq t\}$ do not change under rescaling of the position $x$ with $v$ ( see Eq.~\eqref{RTP-motion}). Finally, observe that in the $\gamma \to \infty$ limit all the three distributions approach to the distribution $P_{Br}(t_a,t)$ obtained for a Brownian motion.  This can be seen using the large argument asymptotic of the Bessel functions 
$I_\nu(z)|_{z \to \infty} \sim \frac{e^{z}}{\sqrt{2 \pi z}}(1+O(z^{-1}))$ for all $\nu$. 

The results in Eq.~\eqref{results} are valid for symmetric initial condition case where the initial values of $\sigma$ (velocity direction) are chosen with equal probability. However, for asymmetric initial conditions in which the directions $+$ and $-$ are chosen with probabilities $a$ and $b$, respectively, such that $a+b=1$, one can straightforwardly extend our calculations and obtain the distributions $P_M(t_m,t)$, $P_R(t_r,t)$ and $P_L(t_\ell,t)$. Explicit expressions and derivations of these probabilities are given in \ref{P_M(t_m,t)_assy} and \ref{P(t_r,t)_assy}.

In addition to the above results, we show (again for $a=b=1/2$) that the joint distribution of  $M$ and $t_m$ is given by 
\begin{align}
\mathcal{P}(M,t_m,t)=& \frac{h(t)}{2} \delta(M)\delta(t_m) -\frac{\delta(t_m-t) + h(t-t_m)}{2} \frac{d}{dM}\left[ \Theta(vt-M)~g(M,t)\right], \label{P_joint-int} \\
  \qquad \text{where}, ~
 g(M,t) &=e^{-\gamma t}\left[I_0\left(\frac{\gamma}{v}\sqrt{v^2t^2 -M^2}\right)+\sqrt{\frac{v t-M}{ vt+M}} I_1\left(\frac{\gamma}{v}\sqrt{v^2t^2 -M^2}\right) \right], \nonumber 
\end{align}
and $h(t)$ is given in Eq.~\eqref{h(t)}. From this joint distribution, we obtain the marginal distribution $\mathcal{P}_M(M,t)$ of the maximum displacement $M$ in time $t$ in addition to the marginal distribution $P_M(t_m,t)$ mentioned before. We show that the cumulative probability 
$Q(M,t)= \int_0^MdM'~\mathcal{P}_M(M',t)$ is given by 
\begin{align}
Q(M,t)=1-\frac{1}{2}g(M,t)\Theta(vt-M) -\frac{\gamma}{2}\int_0^tdt'~\Theta(vt'-M)~h(t-t')~g(M,t').
\end{align}
To prove these results, an essential quantity is the propagator of the RTP in presence of  an absorbing wall for arbitrary initial and final conditions. We start our calculations by computing these propagators in sec.~\ref{prop-abBC}, which is followed by the computation of  the corresponding survival probabilities in the section \ref{surv-RTP}. The results obtained in these sections are used in the later sections to obtain the distributions of $t_m$, $t_r$ and $t_\ell$. In sec.~\ref{Jpdf-M-t_m}, we study at the joint distribution of $M$ and $t_m$ and in sec.~\ref{t_m} we study the marginal distribution of $t_m$. In the last two sections \ref{disr-t_r} and \ref{dist-t_l}, we obtain the distributions of $t_r$ and $t_\ell$ respectively. Finally, in sec.~\ref{conclusion} we conclude the paper. Some details of the calculations are put in the appendices.

\section{Propagator of a RTP in presence of an absorbing barrier at $x=M$}
\label{prop-abBC}
We start by computing the propagators of the RTP in presence of an absorbing boundary. 
Let $P_\pm(x,t|0,\alpha,0)$ represent the probability densities that the RTP, starting at the origin with velocity direction $\alpha \in \{1,-1\}$, reaches the position $x$ at time $t$ with velocity direction $\pm 1$  in presence of an absorbing barrier at $x=M>0$. This particular choice of the position of the absorbing barrier will be useful in later sections where we compute the joint distribution of the maximum displacement and the time at which the maximum occurs. Starting from the Langevin equation Eq.~\eqref{RTP-motion}, it is easy to show \cite{Kanaya18, Doussal19} that these propagators satisfy the following forward master equations 
\begin{align}
\begin{split}
  \partial_t P_+(x,t|0,\alpha,0) &= -v \partial_x P_+(x,t|0,\alpha,0) -\gamma P_+(x,t|0,\alpha,0) + \gamma P_-(x,t|0,\alpha,0), \\
   \partial_t P_-(x,t|0,\alpha,0) &=~~~v \partial_x P_-(x,t|0,\alpha,0) +\gamma P_+(x,t|0,\alpha,0) - \gamma P _-(x,t|0,\alpha,0),
 \end{split}
\label{P_pm}
\end{align} 
with boundary conditions
$P_\pm(x \to -\infty,t|0,\pm,0)=0,~~\text{and}~~P_-(x\to M^-,t|0,\pm,0)=0$
and, the initial conditions $P_+(x,0|0,\alpha,0)=a~\delta_{\alpha,+}\delta(x)$ and $P_-(x,0|0,\alpha,0)=b~\delta_{\alpha,-}\delta(x)$ such that $a+b=1$. The master equations in Eq.~\eqref{P_pm} are also known as Telegrpher's equations and have been studied extensively in various contexts like electromagnetic transmission, chromatography, fluorescence spectroscopy etc. (See \cite{Weiss02} for a review).  
Different choices for the values of $a$ and $b$ correspond to different initial velocity distributions. For example, $a=1$ and $b=0$ implies  that the particle starts with velocity $+v$ initially while $a=0$ and $b=1$ means that the particle starts with velocity $-v$ initially. The boundary conditions stated above can be understood as follows. For finite $t$, the particle, starting from the origin, can never reach $x\to -\infty$ in finite duration $t$ which implies $P_\pm(x \to -\infty,t|0,\pm,0)=0$. 
To see how the boundary condition at $x=M$ appear, we follow \cite{Doussal19} where the problem of finding the propagator of a single RTP in presence of an absorbing boundary (at the origin) has recently been considered. In this paper, the final result of the total probability 
$P(x,t|0,0)=a[P_+(x,t|0,+,0)+P_-(x,t|0,+,0)]+b[P_+(x,t|0,-,0)+P_-(x,t|0,-,0)]$ at the wall has been presented explicitly. We require expressions of the individual probabilities $P_\pm(x,t|0,+,0),~P_\pm(x,t|0,-,0)$ near the absorbing wall to compute the joint distribution of the maximum displacement $M$ and $t_m$,  for which we feel it is convenient to re-derive the solutions of the master equation \eqref{P_pm} in our setting \emph{i.e.} in presence of an absorbing boundary at $x=M$. 

Coming back to the boundary condition at $x=M$, let us write the evolution of the second equation in Eq.~\eqref{P_pm} from time $t$ to $t+dt$ at $x=M$ as $P_-(M,t+dt|0,\alpha,0)=(1-\gamma dt)P_-(M+vdt,t|0,\alpha,0)+\gamma dt ~P_+(M,t|0,\alpha,0)$. Since there can not be any particle at $x>M$ for $t>0$ due to the presence of an absorbing boundary at $x=M$, we have  $P_-(M+vdt,t|0,\alpha,0)=0$. Now taking $dt \to 0$ limit, we have $P_-(M,t|0,\alpha,0)=0$. So finally  the boundary conditions with which the Eqs.~\eqref{P_pm} have to be solved are $P_\pm(x \to -\infty,t|0,\pm,0)=0$ at $x \to -\infty$ and  $P_-(M,t|0,\alpha,0)=0$ at $x=M$. It turns out that these boundary conditions are enough to find the propagator $P_\pm(x,t|0,\alpha,0)$ uniquely\cite{Doussal19}.

\begin{figure}[h!]
\includegraphics[scale=0.24]{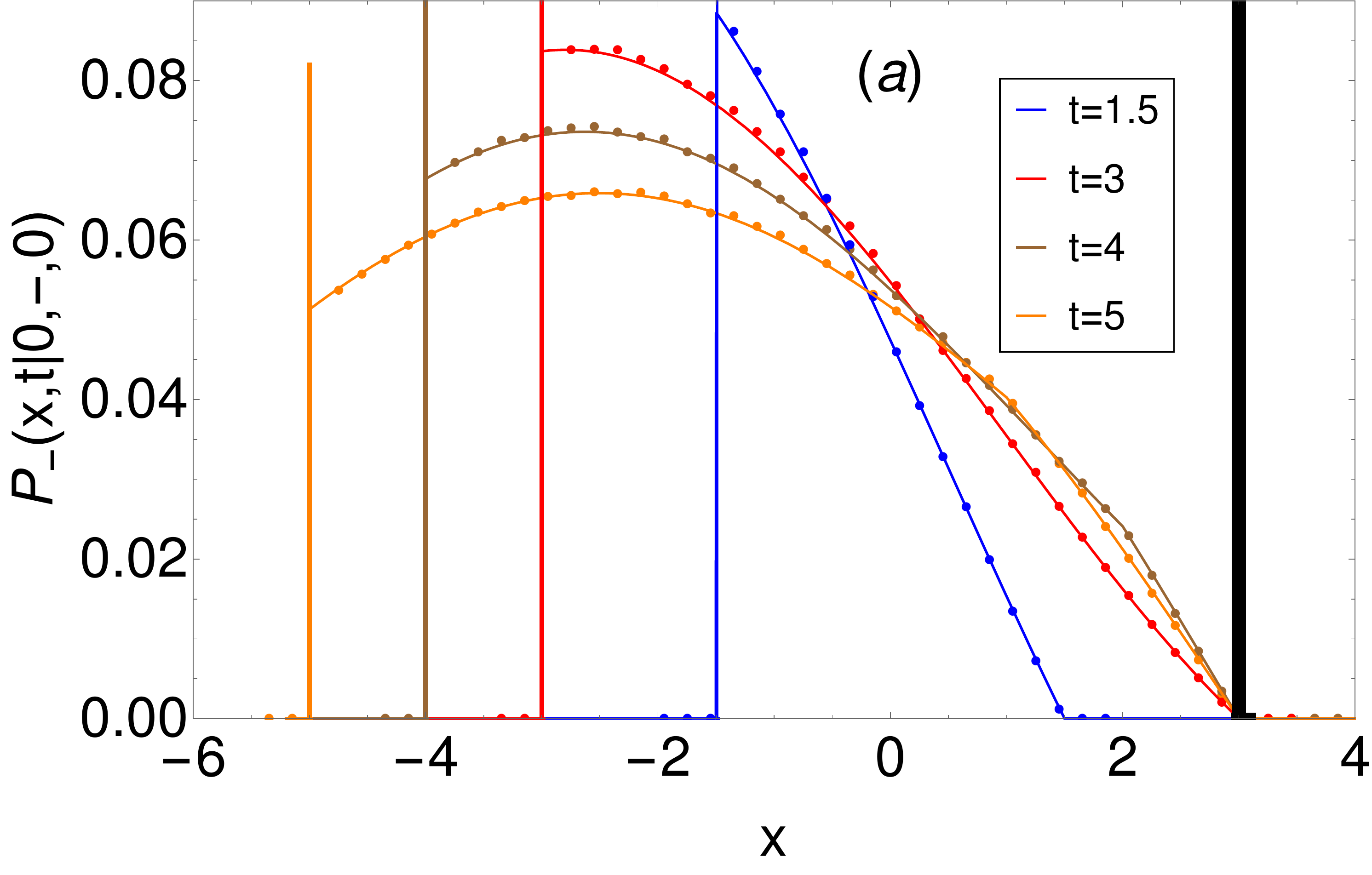}
\includegraphics[scale=0.26]{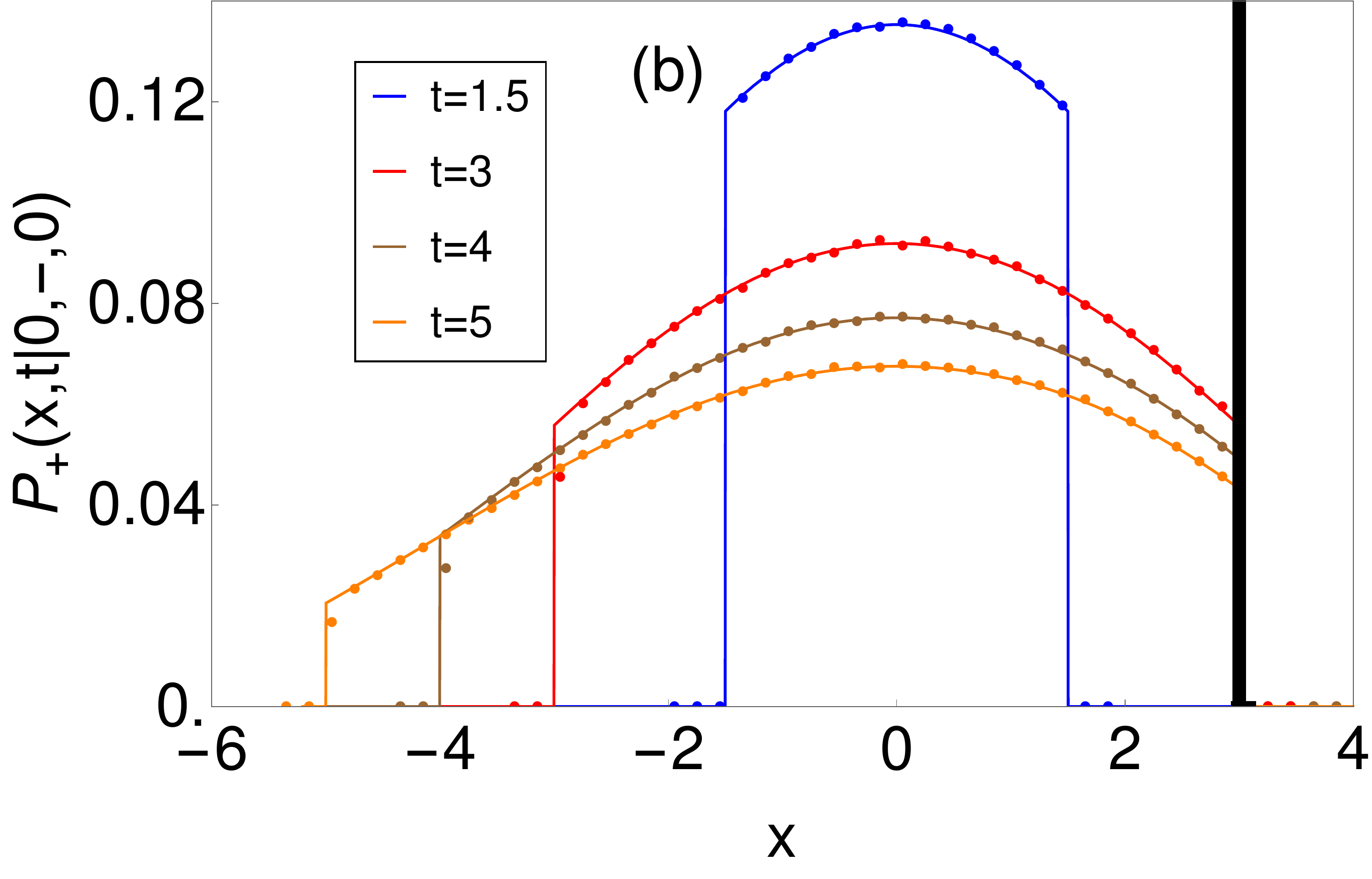}
\centering
\caption{ Numerical verification (filled circles) of the time evolution of the propagators $P_-(x,t|0,-,0)$ and $P_+(x,t|0,-,0)$ given in  Eqs.~\eqref{Pm_m} and \eqref{Pp_m} respectively. We have chosen the absorbing barrier at position $M=3$ (shown by the thick black line). The delta functions The delta functions present in $P_-(x,t|0,-,0)$ at different times are not shown completely. Other parameters of the plot are $v=1$ and $\gamma=0.5$.}
\label{ipm}
\end{figure}

For clarity and compactness of the presentation, we prefer to give the details of the derivation of the propagators in the  \ref{app-propagator}. We here instead present only the final results. We find the following explicit expressions for the propagators for the four possible combinations of the initial and final velocity directions,
\begin{align}
  P_-(x,t|0,+,0)&=\frac{\gamma}{2v}[\mathcal{I}(|x|,v,\gamma,t)
  -\mathcal{I}(2M-x,v,\gamma,t)], \label{Pm_p} \\
P_+(x,t|0,+,0)&=\frac{\gamma}{2v}[\mathcal{J}(|x|,v,\gamma,t)-\mathcal{J}(2M-x,v,\gamma,t)] -\Theta(x)~\frac{d \mathcal{I}(x,v,\gamma,t)}{dx}
\label{Pp_p} \\
P_-(x,t|0,-,0)&=\frac{\gamma}{2v}[\mathcal{J}(|x|,v,\gamma,t)-\mathcal{J}(2M-x,v,\gamma,t)] -\Theta(-x)~\frac{d \mathcal{I}(|x|,v,\gamma,t)}{d|x|}
    ~~\text{and}
\label{Pm_m} \\
P_+(x,t|0,-,0)&=\frac{\gamma}{2v}[\mathcal{I}(|x|,v,\gamma,t)-\mathcal{I}(2M-x,v,\gamma,t)] + \mathcal{T}(2M-x,v,\gamma,t)
\label{Pp_m} 
\end{align}
where the functions $\mathcal{I}(g,v, \gamma,t)$, $\mathcal{J}(\alpha,g,v,\gamma,t)$ and $\mathcal{T}(g,v,\gamma, t)$ (defined for $g>0$) are given by:
\begin{align}
  \mathcal{I}(g,v, \gamma,t)&= \Theta \left(v t-g\right)e^{-\gamma t} I_0\left( \gamma \frac{\sqrt{ v^2t^2-g^2}}{v}\right),
\label{I} \\
\mathcal{J}(g,v,\gamma,t)&= \Theta \left(vt-g \right) e^{-\gamma t}  \sqrt{\frac{vt-g}{vt+g}} I_1\left(\gamma \frac{\sqrt{v^2 t^2-g^2}}{v} \right),
\label{JJ} \\
\begin{split}
\mathcal{T}(g,v,\gamma, t)&= \Theta \left(vt-g\right) \frac{ e^{-\gamma t}}{v t+g}\left[\frac{\gamma g}{v} I_0\left( \frac{\gamma\sqrt{v^2 t^2-g^2}}{v}\right) \right. \\
&~~~~~~~~~~~~~~~~~~~~~~~~~~~~~~~~~~\left.+\sqrt{\frac{v t-g}{v t+g}} I_1\left( \frac{\gamma\sqrt{v^2 t^2-g^2}}{v}\right)\right] .
\end{split}
\label{TT}
\end{align}
\begin{figure}[t] 
  \includegraphics[scale=0.3]{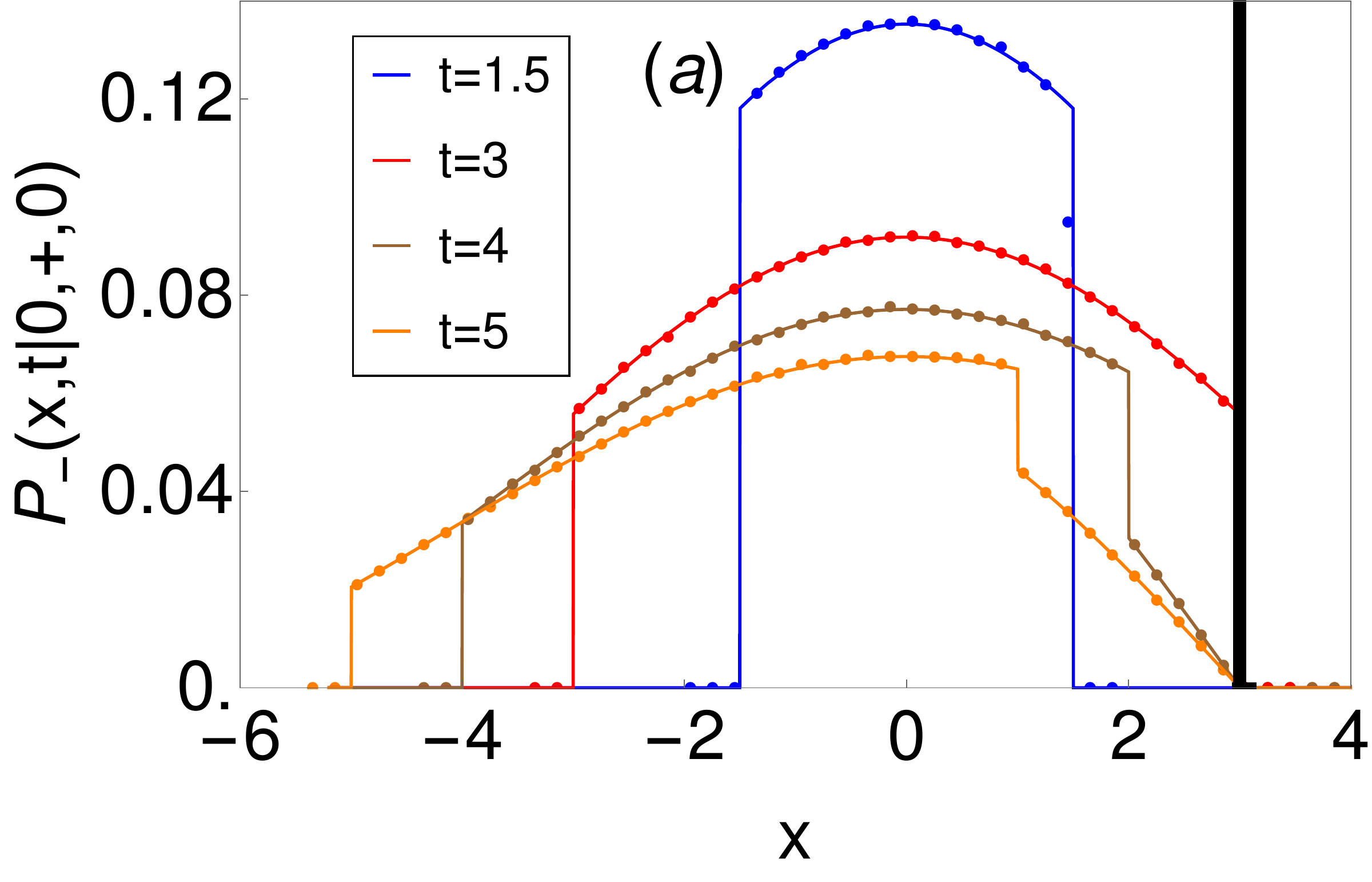}  
   \includegraphics[scale=0.3]{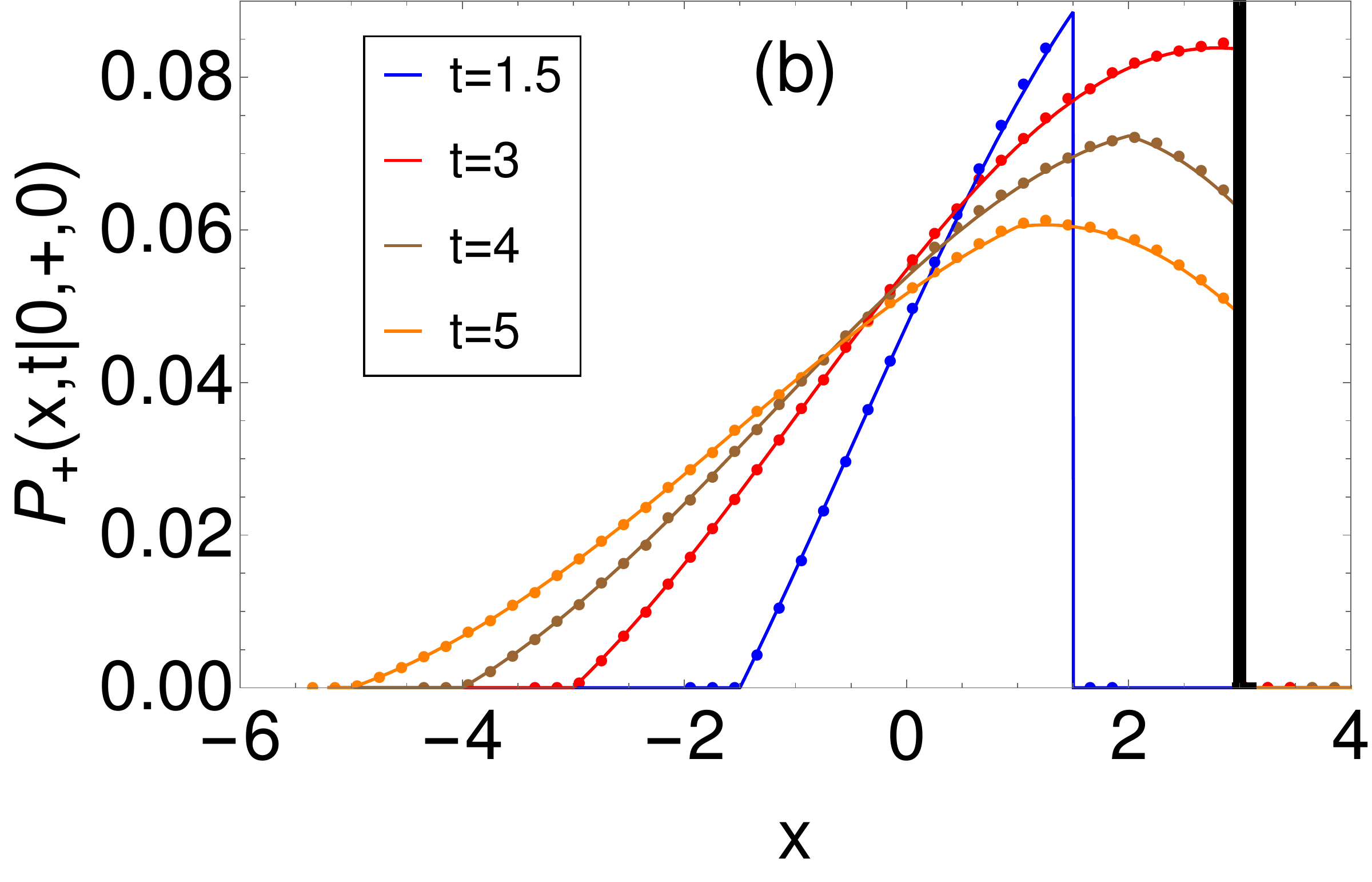}
   \centering
\caption{Numerical verification (filled circles) of the time evolution of the propagators  $P_-(x,t|0,+,0)$ and $P_+(x,t|0,+,0)$ given in  Eqs.~\eqref{Pm_p} and \eqref{Pp_p} respectively. We have chosen the absorbing barrier at position $M=3$ (shown by the thick black line). The delta function present in $P_+(x,t|0,+,0)$ is not shown in the plot. Other parameters of the plot are $v=1$ and $\gamma=0.5$. }
\label{ipp}
\end{figure}
\noindent
Observe that the term $\frac{d \mathcal{I}}{dx}$ would produce some delta function terms like $e^{-\gamma t}\delta(x \pm vt)$. 
These terms arise from the events in which the RTP has not changed its direction till time $t$ and such events occur with probability $e^{-\gamma t}$. 
From these expressions, we can obtain the total probability density $P(x,t|x,\alpha,0)=a[P_+(x,t|0,+,0)+P_-(x,t|0,+,0)]+b[P_+(x,t|0,-,0)+P_-(x,t|0,-,0)]$ of finding the particle at $x$ in time $t$ and check that at $x=M$ it matches with the result given in \cite{Doussal19}.
Using the large $z$ asymptotic of $I_{\nu}(z) \simeq  e^z/\sqrt{2\pi z}$ for both $\nu=0$ and $1$, one can show that in the limit $v \to \infty$ and $\gamma \to \infty$ keeping $v^2/(2 \gamma)=D$ fixed, the individual probability densities in Eqs.~\eqref{Pm_p} - \eqref{Pp_m} reduce to 
\begin{align}
P_\alpha(x,t|0,\alpha',0) \xrightarrow[\gamma \to \infty,~v \to \infty ]{ ({v^2}/{2\gamma}) =D}  \frac{1}{\sqrt{4 \pi D t}}\left[ \exp\left( -\frac{x^2}{4 D t}\right)-\exp\left( -\frac{(2M-x)^2}{4 D t}\right)\right],
\end{align}
for $\alpha=\pm 1$, $\alpha'=\pm1$ and for $x<M$. This is the propagator of a Brownian walker in presence of absorbing boundary at the origin, with the initial and final positions are $y_0=M$ and $y_t=M-x$, respectively. 
Hence the total probability $P(x,t|0,0)=a[P_+(x,t|0,+,0)+P_-(x,t|0,+,0)]+b[P_+(x,t|0,-,0)+P_-(x,t|0,-,0)]$ with $a+b=1$ for $x<M$, reduces to the Brownian motion result with an absorbing boundary at $x=M$, as it should. One obtains the same Brownian motion result in the large $t$ asymptotic as well. On the other hand, for finite $\gamma$,  the effect of the activity will be stronger and to observe these effects, we plot the probability densities in Eqs.~\eqref{Pm_p} - \eqref{Pp_m} as functions of $x$ for $\gamma=0.5$ at different times in Figs.~\ref{ipp} and \ref{ipm}. In these figures we observe excellent agreement with the simulation data (filled circles). 

\section{Survival probability of the RTP in presence of the absorbing barrier at $x=M$}
\label{surv-RTP}
\noindent
Here we look at the survival probability of the RTP in presence of the absorbing barrier at $x=M>0$. In principle  one can compute the survival probabilities by integrating out  final positions over $(-\infty,M)$ in the propagators  obtained in the previous section. However, it seems more convenient and illustrative to solve the backward master equation to get the survival probabilities. The survival probability with an absorbing barrier at $x=0$ for a run-and-tumble particle has recently been studied in \cite{Kanaya18}. We mainly use results from \cite{Kanaya18} after appropriately modifying for our case.  However, for self-containedness of the paper we here briefly outline the derivation of the survival probabilities. At this point we also would like to mention that a generalisation of the  survival probability problem has been recently addressed in the context of RTP with resetting \cite{Evans19}.

Let $S_\pm(x,t)$ represent the survival probability of the RTP, starting from the origin with velocity direction $\pm1$.    Note that in our problem, the absorbing barrier is at $M\neq 0$ and the particle starts at $x\leq M$. The backward master equations for the survival probability $S_{\pm}(x, t)$ are given by \cite{Kanaya18}
\begin{align}
 \partial_t S_+(x,t) &= v \partial_x S_+(x,t) -\gamma S_+(x,t) + \gamma S_-(x,t), \label{S+d} \\
 \partial_t S_-(x,t) &=- v \partial_x S_-(x,t) +\gamma S_+(x,t) - \gamma S_-(x,t). \label{S-d}
\end{align}
We solve the above equations with initial condition $S_{\pm}(x,0;M)=1$ and the boundary conditions $S_{\pm}(x\to -\infty,t;M)=1$ and $S_+(M^-,t;M)=0$. The first boundary condition takes into account of the fact that  the particle, regardless of its initial velocity direction,  will always survive over any finite time if it starts initially from $x\to -\infty$. On the other hand if the particle starts at $x=M$ with positive velocity then the particle will immediately get absorbed.  With these conditions one can solve Eqs.~\eqref{S+d} and \eqref{S-d}  to find \cite{Kanaya18}
\begin{align}
S_+(x,t)&=1+v\frac{d}{dM}\int^t_0 d \tau~ {\Theta\left(\tau-\frac{M-x}{v}\right)}  e^{-\gamma \tau}~ I_0
 \left(\frac{\gamma}{v}\sqrt{v^2\tau^2 -(M-x)^2}\right) ,   
 \label{Sp}  \\
S_-(x,t) &= 1- \int_0^t d\tau ~{\Theta\left(\tau-\frac{M-x}{v}\right)}  \frac{v e^{-\gamma \tau}}{ v\tau+(M-x)}\left[\frac{\gamma}{v} (M-x)I_0 \left(\frac{\gamma}{v}\sqrt{v^2\tau^2 -(M-x)^2}\right) \right.  \nonumber \\
      &~~~~~~~~~~~~~~~~~~~~
      \left. +\sqrt{\frac{v \tau-(M-x)}{ v\tau+(M-x)}} I_1\left(\frac{\gamma}{v}\sqrt{v^2\tau^2 -(M-x)^2}\right)\right ].
\label{Sm}
\end{align}
\begin{figure}[t]
\includegraphics[scale=0.36]{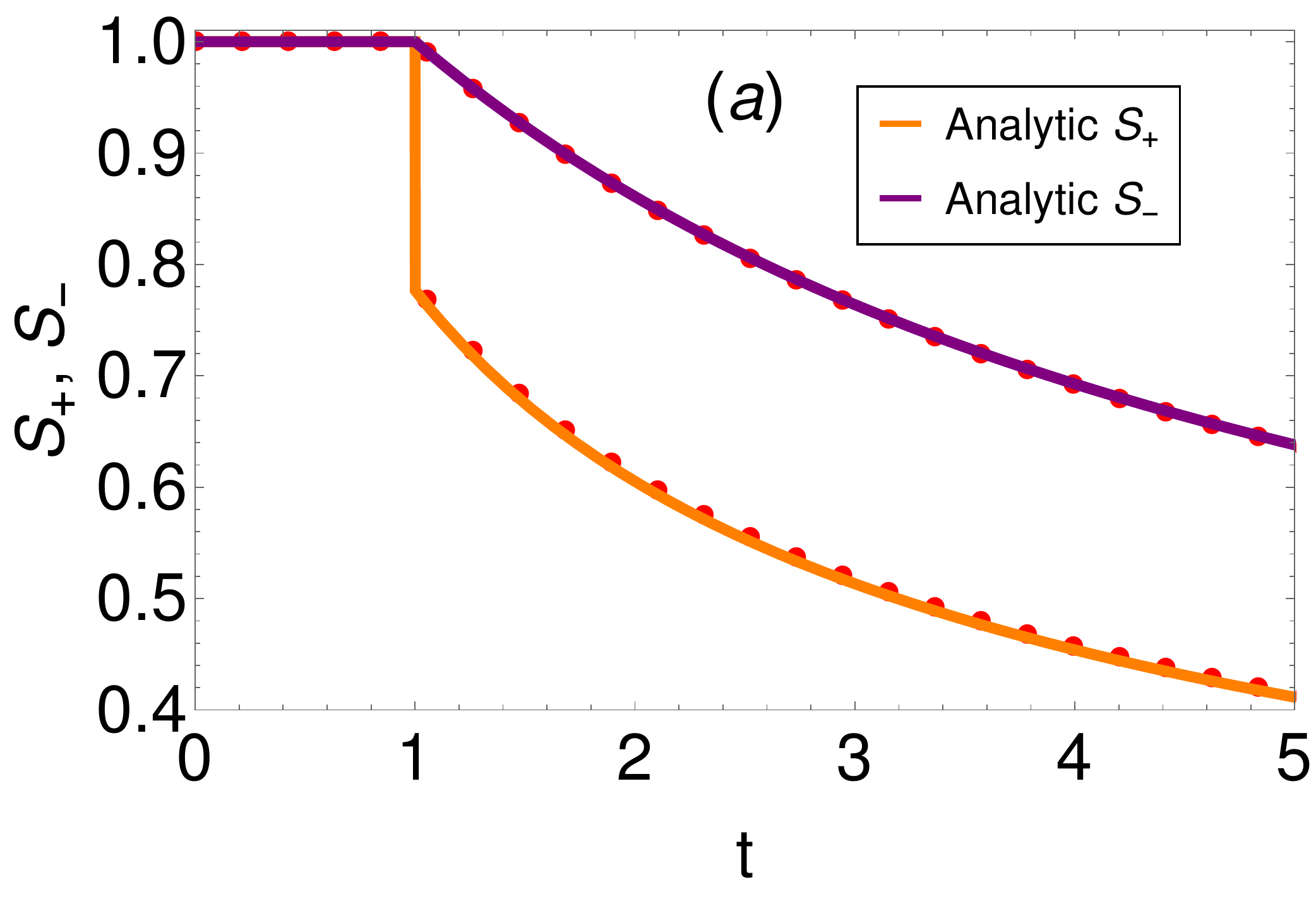}
\includegraphics[scale=0.41]{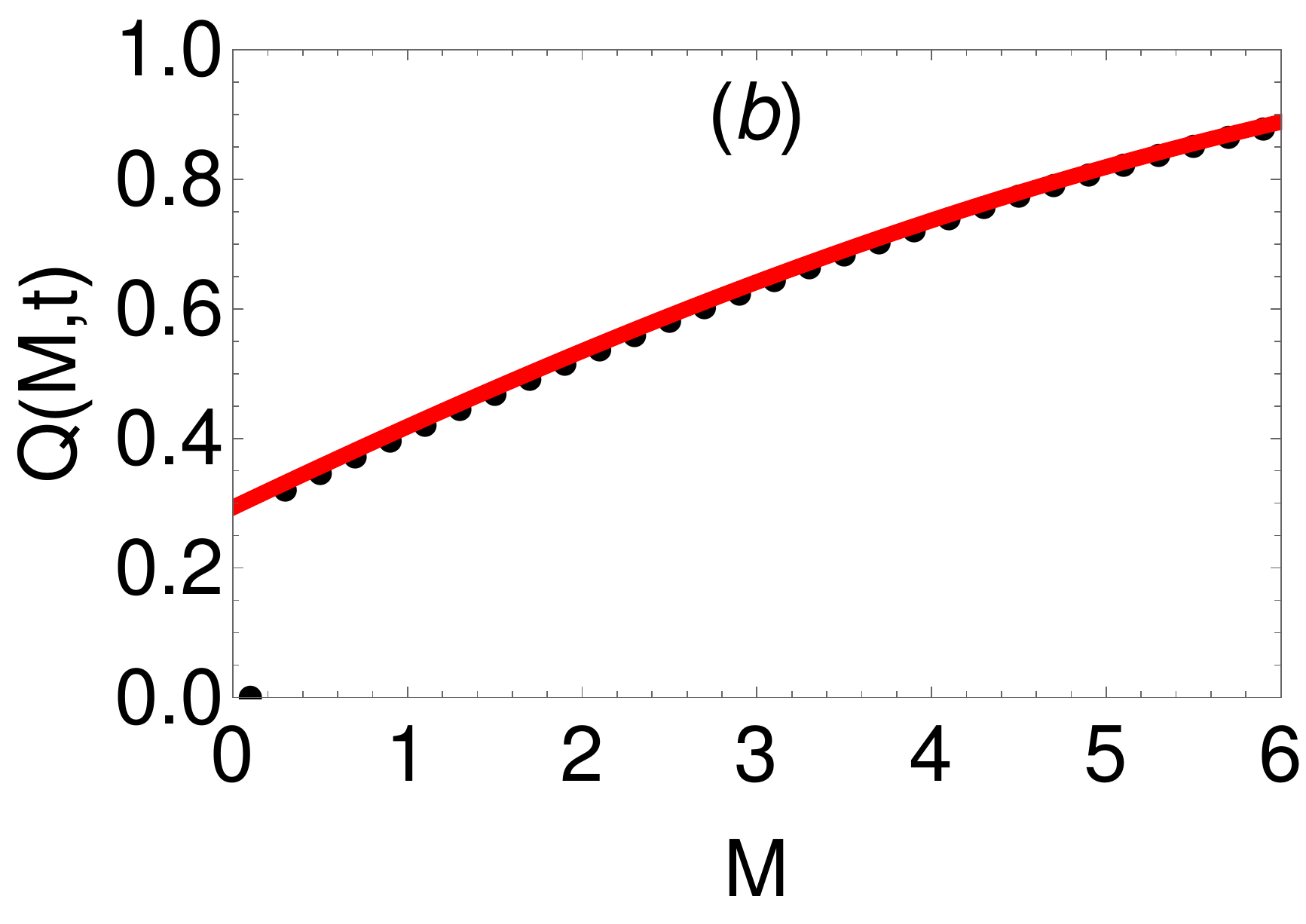}
\centering
\caption{(a) Comparison of the survival probabilities $S_\pm$ given in Eqs.(\ref{Sp}) and (\ref{Sm}) with simulation results (filled circles). We have chosen the initial position at $x=1$ and the absorbing barrier at $M=3$.  (b) Numerical verification of analytical result of the cumulative distribution $Q(M,t)=[S_+(0,t)+S_-(0,t)]/2$ of the maximum displacement $M$ made by the particle in time $t=3$, starting from the origin. The discontinuous jumps at $M=0$ and $M=vt$ are due to the presence of delta functions. Other parameters used in both the plots are $\gamma=2.5, v=2$.}
\label{survpic1}
\end{figure}
The results in Eqs.~\eqref{Sp} and \eqref{Sm} are plotted in Fig.~\ref{survpic1} and compared with the numerical simulations, in which one starts with a large number of independent RTPs intially. We observe excellent agreement. We notice that both $S_\pm $ remains $1$ till the time $ t_b=\frac{M-x}{v}$. This is because the RTPs, starting from $x<M$, reach the absorbing barrier for the first time 
at least after time $t_b$. Before this time the RTPs do not feel the presence of the absorbing barrier.  Once the RTPs reach (they can reach only with $+$ve velocity) the barrier at $t_b$, in the next instant some fraction of the RTP change their velocity direction from $+$ to $-$ and the rest of the RTPs do not change. The fraction of particles which do not change their velocity direction get absorbed at the barrier at $t_b^+$ time.  As a result $S_+(x,t_b^+)$ drops suddenly as manifested by a discontinuity in the Fig.~\ref{survpic1}a. No sudden drop occurs $S_-(x,t)$ because no particles with $-$ve velocity can reach the barrier at $t_b^-$. However, since the overall number of particles decrease, the $S_-(x,t)$  also starts decreasing from value $1$ after time $t_b$ but continuously.  As done in the previous section, using the large argument asymptotic behaviour of $I_\nu(z)$, one can show that in the $\gamma \to \infty$ and $v \to \infty$ limit keeping $v^2/(2\gamma)=D$ fixed, one indeed recovers the survival probability of a Brownian particle 
from the absorbing wall at $M$ when started from $x$:
\begin{align}
S_\pm(x,t) \xrightarrow[\gamma \to \infty,~v \to \infty ]{  ({v^2}/{2\gamma}) =D} \text{erf}\left( \frac{M-x}{\sqrt{4 D t}} \right)~~\text{where,~erf}(z)=\frac{2}{\sqrt{\pi}}\int_0^zdu~e^{-u^2}.
\label{S-Brow}
\end{align}

\subsection{Cumulative probability $Q(M,t)$ of maximum displacement $M$ in absence of any absorbing barrier}
\label{maximum}
\noindent
It is easy to see that the distribution of the maximum $M$ over the duration $t$ in absence of any absorbing barrier is related to the survival probability of the RTP in presence of an absorbing barrier at $M$. The cumulative probability $Q(M,t)$ of the maximum displacement made by the RTP in time duration  $t$ (given that it started from the origin) is exactly same as the survival probability of the particle till time $t$ in presence of an absorbing barrier at $M$ and starting from $x=0$.   In both the probabilities, exactly the same ensemble of paths contribute which, starting from $x=0$ stays below the level $x=M$ throughout the course of evolution till time $t$. For a discussion on the connection between $Q(M,t)$ and survival probability in the context of Brownian motion see \cite{Majumdar10-rev}. This connection allows us to write, 
\begin{align}
Q(M,t)&=\text{Prob.}(x_{\text{max}} \leq M,t)=[S_+(0,t)+S_-(0,t)]/2,~~~\text{where} \label{Q(M,t)} \\
 \begin{split}
S_+(0,t)&=1+v\frac{d}{dM}\int^t_0 d \tau ~{\Theta\left(v\tau-M\right)} e^{-\gamma \tau} I_0
 \left(\frac{\gamma}{v}\sqrt{v^2\tau^2 -M^2}\right),   \\
S_-(0,t) &= 1-  \int_0^t d\tau~{\Theta\left(v\tau-M\right)} \frac{v e^{-\gamma \tau}}{ v\tau+M}\left[\frac{\gamma M}{v} I_0 \left(\frac{\gamma}{v}\sqrt{v^2\tau^2 -M^2}\right)   \right. \\
 &~~~~~~~~~~~~~~~~~~~~~~~~~~~~~~~
 \left.+\sqrt{\frac{v \tau-M}{ v\tau+M}} I_1\left(\frac{\gamma}{v}\sqrt{v^2\tau^2 -M^2}\right)\right ],
\end{split}
\label{SQm}
\end{align}
where the factor $1/2$ is due to the choice $a=b=1/2$. These expressions agree with the results obtained in \cite{Masoliver93} where the maximum displacement made by a persistent Brownian walker was computed. 
From Eq.~\eqref{S-Brow}, one can easily see that in the diffusive asymptotic limit one correctly recovers the 
Brownian result. In Fig.~\ref{survpic1}b, we compare our theoretical cumulative distribution $Q(M,t)$ against numerical simulation and we observe nice agreement. The discontinuous jumps at $M=0$ and $M=vt$ indicate the presence of delta functions in the corresponding probability distribution function $\mathcal{P}_M(M,t)$ which we will see in the next section.

\begin{figure}[t]
 \includegraphics[scale=0.5]{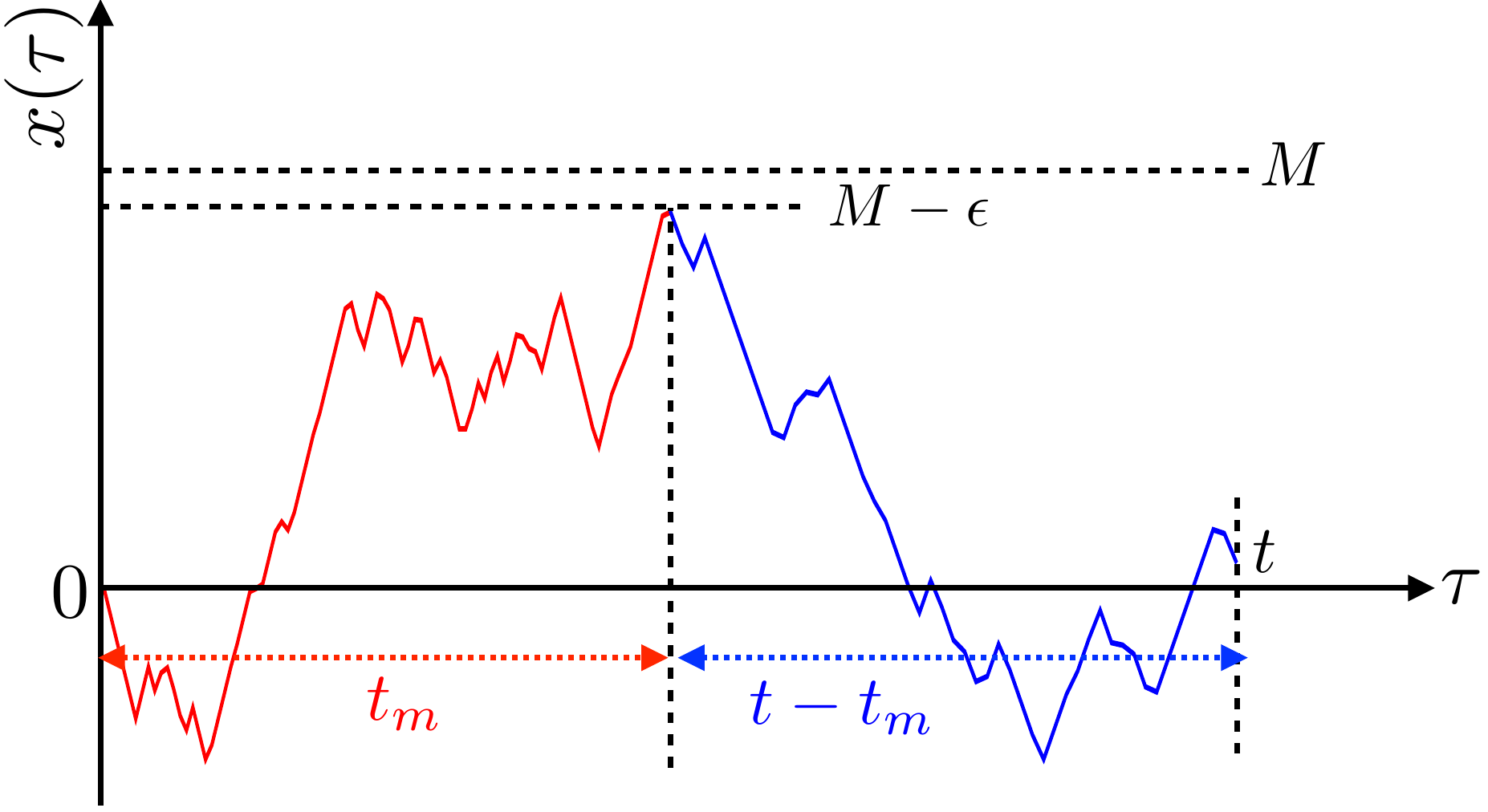}
\centering
\caption{A typical trajectory starting from the origin reaches the position $M-\epsilon$ at time $t_m$ and in the remaining time $t-t_m$ reaches some position $x_f$ in presence of an absorbing boundary at $x=M$. The trajectory is decomposed into two parts, from $0$ to $t_m$ (red part) and from $t_m$ to $t$ (blue part). } 
\label{traj-M-ep}
\end{figure}

\section{Joint probability distribution $\mathcal{P}(M,t_m,t)$ of $M$ and $t_m$ and the marginal distributions}
\label{Jpdf-M-t_m}
Let us now focus on the joint distribution $\mathcal{P}(M,t_m,t)$ of the maximum $M$ and the time $t_m$ at which this maximum occurs within the time interval  $[0,t]$.  To compute this distribution, we first  break the trajectory $x(\tau)$ into two parts (see Fig.~\ref{traj-M-ep}). The first part is $\{x(\tau); 0\leq \tau \leq t_m\}$  and the second part is $\{x(\tau); t_m\leq \tau \leq t\}$. In the first part $0\leq \tau \leq t_m$, the trajectory of the RTP is always below $M$ \emph{i.e.} $x(\tau) <M$ and reaches $M$ exactly at $t_m$ (with velocity $+v$ ). Similarly, in the second part, the trajectory, starting from $M$ also stays below $M$ throughout in the remaining duration $t_m < \tau \leq t$, however, the final position can be anywhere below $M$.  The probability weight in the first part can be obtained from the propagator with an absorbing barrier at $M$. Since in the variables $(x,\sigma)$, the RTP process $\{x(\tau),\sigma(\tau)\}$ is Markovian, the probability weight of the paths in the second part is independent of that of the first part and is proportional to the survival probability of the RTP starting from $M$ in presence of a barrier at $M$. It is intuitively clear that in the first interval of time, the RTP can not reach  $M$ (at time $t_m$) with negative velocity while  staying completely below $M$ over the full duration $0\leq \tau \leq t_m$.  Similarly, in the second interval, the RTP can not  survive  from the absorbing wall at $M$  if it starts with a positive velocity at $x=M$.  From the expressions of the propagators in Eqs.~(\ref{Pm_p}-\ref{Pp_m}) and the survival probabilities in Eqs~(\ref{Sp}) and \eqref{Sm}, it is easy to observe that if one puts $x=M$ then $P_-(M,t_m|0,\pm,0)=0$ and $S_+(M,t)=0$. To proceed at this point, we follow the procedure given in \cite{ Majumdar08}. We use the propagators $P_\pm(M-\epsilon,t_m|0,0)$ that the RTP, starting at the origin, reach $M-\epsilon$ (for $\epsilon >0$) with velocity $\pm v$ at time $t_m$ and, the probabilities $S_\pm(M-\epsilon,t-t_m)$ that the RTP, starting from position $M-\epsilon$ with velocity $\pm v$ survives over the duration $t-t_m$. Later at some appropriate step, we take the $\epsilon \to 0$ to get the final answers. In terms of these propagators and survival probabilities, we can write  
\begin{equation}
\mathcal{P}(M,t_m,t) = \lim_{\epsilon \to 0} \left[\frac{p_+(M-\epsilon, t_m|0,0) S_+(M-\epsilon, t-t_m) 
+ p_-(M-\epsilon, t_m|0,0)S_-(M-\epsilon, t-t_m)}{\mathcal{Z}(\epsilon,t)}\right],
\label{JPdf-nd}      
\end{equation}
for $0<M<vt$ and $0<t_m<t$ where $\mathcal{Z}(\epsilon, t)$ is a normalization factor and  the propagators $p_\pm(M-\epsilon,t_m|0,0)$ each are the sum of two terms corresponding to two initial velocities ( occurring with probability $1/2$ each) \emph{i.e.} $p_\pm(M-\epsilon,t_m|0,0)=\frac{1}{2} \left[P_\pm(M-\epsilon, t_m | 0, +, 0)+P_\pm(M-\epsilon, t_m | 0, -, 0) \right].$ We emphasise that the propagators in Eqs.~\eqref{Pm_p}-\eqref{Pp_m} were derived for biased initial conditions with $(a, b)=(1,0) \text{ or } (0,1)$. However while writing $p_{\pm}$, we consider $a=b=\frac{1}{2}$.
To take the $\epsilon \to 0$ limit, we first expand each terms separately in $\epsilon$. We find 
\begin{align}
& \left.\begin{aligned}
S_-\left(M-\epsilon,t-t_m\right) & \approx \Theta \left( t-t_m \right)  h(t-t_m) + O\left(\epsilon \right),     \\
S_+(M-\epsilon,t-t_m) &\approx \frac{\gamma~\epsilon}{v} h(t-t_m)+O(\epsilon^2),  
       \end{aligned}
 \right\}
 \qquad h(t)=e^{-\gamma t}[I_0(\gamma t)+I_1(\gamma t)], 
 \label{S_pm-ep} \\
&\left.\begin{aligned} 
p_+(M-\epsilon,t_m|0,0)  &\approx  -\frac{d}{dM}\left[ \Theta(vt_m-M)~g(M,t_m)\right] + O\left(\epsilon \right),     \\
p_-(M-\epsilon,t_m|0,0)   &\approx  -\frac{\gamma~\epsilon}{v} \frac{d}{dM}\left[ \Theta(vt_m-M)g(M,t_m)\right]+O(\epsilon^2),  
       \end{aligned}
 \right\}
 \qquad \text{where}, \label{P_pm_ep}\\
 &g(M,t)=e^{-\gamma t}\left[I_0\left(\frac{\gamma}{v}\sqrt{v^2t^2 -M^2}\right)+\sqrt{\frac{v t-M}{ vt+M}} I_1\left(\frac{\gamma}{v}\sqrt{v^2t^2 -M^2}\right) \right].
 \label{g(M,t)}
\end{align}
Note that, $h(t)$ is the survival probability of the RTP in presence of an absorbing barrier at the origin while starting from the origin itself \cite{Doussal19}.
Inserting the small $\epsilon$ expansions from Eqs.~\eqref{S_pm-ep} and \eqref{P_pm_ep} into Eq.~\eqref{JPdf-nd} one can get the joint distribution of $M$ and $t_m$. Note, however, that  the above procedure can provide the distribution only for $0<M<vt$ and $0<t_m<t$. The contributions from events $M=0$ (or equivalently $t_m=0$) and $t_m=t$ have to be added separately. Contribution to the $M=0$ (or $t_m=0$) comes from trajectories which  starting from the origin goes to the negative $x$ and never visits the positive $x$ region within time $t$.   Probability with which such events occur is exactly the probability  that the RTP, starting from the origin, survives from an absorbing barrier at the origin till time $t$. This survival probability can be easily computed similarly as done in sec.~\ref{surv-RTP} and is given exactly by $h(t)$ ( see also \cite{Doussal19}). On the other hand the contributions to $t_m=t$ event come from trajectories which starting from the origin reach the maximum displacement $M$ at time $t_m=t$. The contribution of these paths to the joint distribution $\mathcal{P}(M,t_m,t)$ is given by
\begin{align}
\underbrace{\mathcal{P}(M,t_m,t)}_{\text{contribution~from~paths~with~}t_m=t} = \delta(t_m-t)~\frac{[P_+(M,t|0,+,0)+P_+(M,t|0,-,0)]}{2}.
\label{pam}
\end{align}
where $\frac{[P_+(M,t|0,+,0)+P_+(M,t|0,-,0)]}{2}$ is the probability density to find the particle at time $t$ in position $M$.  Inserting the explicit forms of these propagators from Eqs.~\eqref{Pp_p} and \eqref{Pp_m}, respectively, we get 
\begin{align}
\underbrace{\mathcal{P}(M,t_m,t)}_{\text{contribution~from~paths~with~}t_m=t} &= \delta(t_m-t)~\frac{1}{2}\left[ \mathcal{T}(M,v,\gamma,t) - \frac{d \mathcal{I}(M,v,\gamma,t)}{dM}\right] \nonumber \\ 
&=- \frac{d}{dM}\left[ \Theta(vt-M)~g(M,t)\right],
\end{align}
where $g(M,t)$ is given in Eq.~\eqref{g(M,t)}.
\begin{figure}[t]
 \includegraphics[scale=0.47]{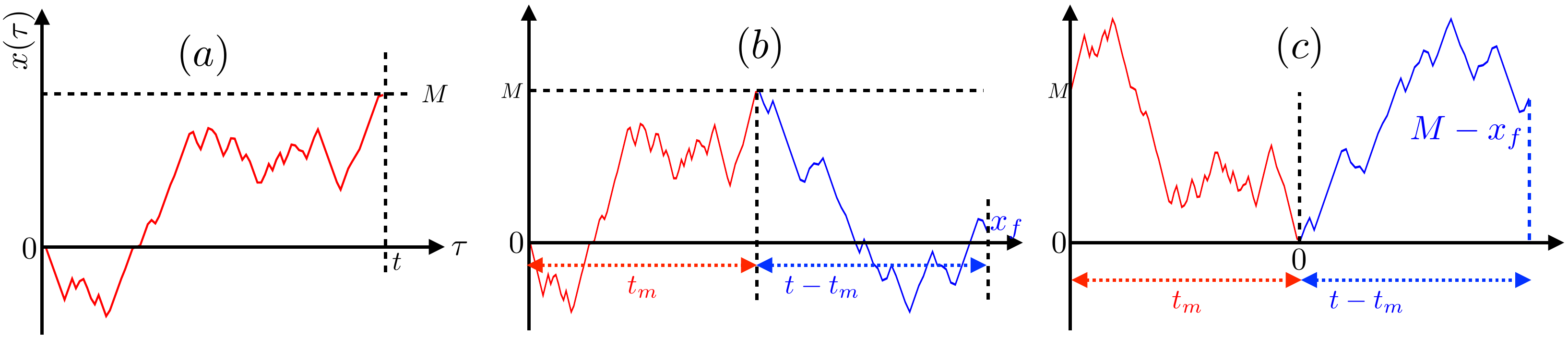}
 \centering
\caption{(a) An example of the trajectories which contribute to the event $t_m=t$ \emph{i.e.} the maximum occurs at time $t$. (b) A typical trajectory which contribute to the non-delta function part of the distribution $\mathcal{P}(M,t_m,t)$. The trajectory is divided into two parts, from $0$ to $t_m$ (red part) and from $t_m$ to $t$ (blue part) (c) Illustration of how the trajectory in (b) look like after the transformations  $\bar{x}=M-x,~\bar{\tau}=t_m-\tau$ in the first (red) part and $\bar{x}=M-x,~\bar{\tau}=\tau-t_m$ in the second (blue) part. } 
\label{traj-inversion}
\end{figure}
Hence, adding all the three contributions, as mentioned above, one gets
\begin{align}
\begin{split}
\mathcal{P}(M,t_m,t)=& \frac{h(t)}{2} \delta(M)\delta(t_m) -\frac{\delta(t_m-t)}{2}\frac{d}{dM}\left[ \Theta(vt-M)~g(M,t)\right] \\
 &-\frac{\gamma \epsilon ~h(t-t_m) }{2v~\mathcal{Z}(\epsilon,t)} \frac{d }{dM} \left[\Theta(vt_m-M)g(M,t_m)\right],
 \end{split}
  \label{P_joint-1}
\end{align}
where $h(t)$ and $g(M,t)$ are given in Eqs.~\eqref{S_pm-ep} and \eqref{g(M,t)}. The factor $1/2$ in the first term comes from the probability with which the RTP, starting from the origin, chooses $-$ve velocity initially. The factor $\mathcal{Z}(\epsilon,t)$ is  a normalisation constant and has to be chosen in such a way that, 
\begin{align}
\int_0^tdt_m \int_0^{vt}dM~\mathcal{P}(M,t_m,t)=1. \label{Norm-1}
\end{align}
The contribution  from the first term in Eq.~\eqref{P_joint-1} to the total unit probability is $h(t)/2$ as can be easily seen
\begin{align}
\begin{split}
&\text{Contribution~from}\\
&\text{the~first~term} 
\end{split}
=\int_0^tdt_m \int_0^{vt}dM \frac{h(t)}{2}\delta(M)\delta(t_m)=  \frac{h(t)}{2}. \label{1st-C}
\end{align}
It is also easy to show  that the contribution from second term (see \ref{cum_dist}) is 
\begin{align}
\begin{split}
&\text{Contribution~from}\\
&\text{the~second~term} 
\end{split}
=
-\int_0^tdt_m \int_0^{vt}dM \frac{\delta(t_m-t)}{2}\frac{d}{dM}\left[ \Theta(vt-M)~g(M,t) \right] =  \frac{h(t)}{2},
 \label{2nd-C} 
\end{align}
where we have used the explicit expression of $g(M,t)$ given in Eq.~\eqref{g(M,t)}. 
Similarly, one can show that the contribution from the third term is 
\begin{align}
\begin{split}
&\text{Contribution~from}\\
&\text{the~third~term} 
\end{split}
=- \frac{\gamma \epsilon  }{2v~\mathcal{Z}(\epsilon,t)}  \int_0^tdt_m h(t-t_m) \int_0^{vt}dM~\frac{d }{dM} \left[\Theta(vt_m-M)g(M,t_m)\right] \nonumber \\
&~~~~~~~~~~~~~~~~~~~~~~~~=\frac{ \epsilon  }{2v~\mathcal{Z}(\epsilon,t)}~2[1-h(t)],  \label{3rd-C} 
\end{align}
where the identity $\gamma\int_0^tdt' h(t-t')h(t')=2[1-h(t)]$ ( see \ref{integration} for the proof) has been used.  Adding all the contributions and substituting in Eq.~\eqref{Norm-1}, we get $\mathcal{Z}(\epsilon,t)=\frac{v}{ \epsilon}$, inserting which in Eq.~\eqref{P_joint-1}, we finally obtain the  joint distribution of $M$ and $t_m$:
\begin{align}
\begin{aligned}
\mathcal{P}(M,t_m,t)=& \frac{h(t)}{2} \delta(M)\delta(t_m) - \frac{\delta(t_m-t) + \gamma~h(t-t_m)}{2} \frac{d\left[ \Theta(vt_m-M)~g(M,t_m)\right]}{dM}. \label{P_joint}
\end{aligned}
\end{align}
This is one of our main results. Although this result is derived for $a=b=1/2$ but for other choices of $a$ and $b$ (such that $a+b=1$), once can easily extend the above calculation (see \ref{P_M(t_m,t)_assy}).

Integrating out the variables $t_m$ or $M$ in the above joint distribution in Eq.~\eqref{P_joint}, one can get the marginal distributions $\mathcal{P}_M(M,t)$ and $P_M(t_m,t)$, respectively. Let us first look at the marginal distribution of the maximum displacement $M$.  It turns out convenient to compute the cumulative distribution $Q(M,t)=\int_0^MdM'~\mathcal{P}_M(M',t)=\int_0^MdM' \int_0^tdt_m~\mathcal{P}(M,t_m,t)$ (See also before Eq.~\eqref{Sp}).   Performing the integrals along with some straight forward manipulations (see \ref{cum_dist}), we find 
\begin{align}
Q(M,t)=1-\frac{1}{2}g(M,t)\Theta(vt-M) -\frac{\gamma}{2}\int_0^tdt'~\Theta(vt'-M)~h(t-t')~g(M,t'),
\end{align}
which can be shown to agree with our earlier result for $Q(M,t)$ given in Eqs.~\eqref{Q(M,t)} and \eqref{SQm}. 

\begin{figure}[t]
 \includegraphics[scale=0.4]{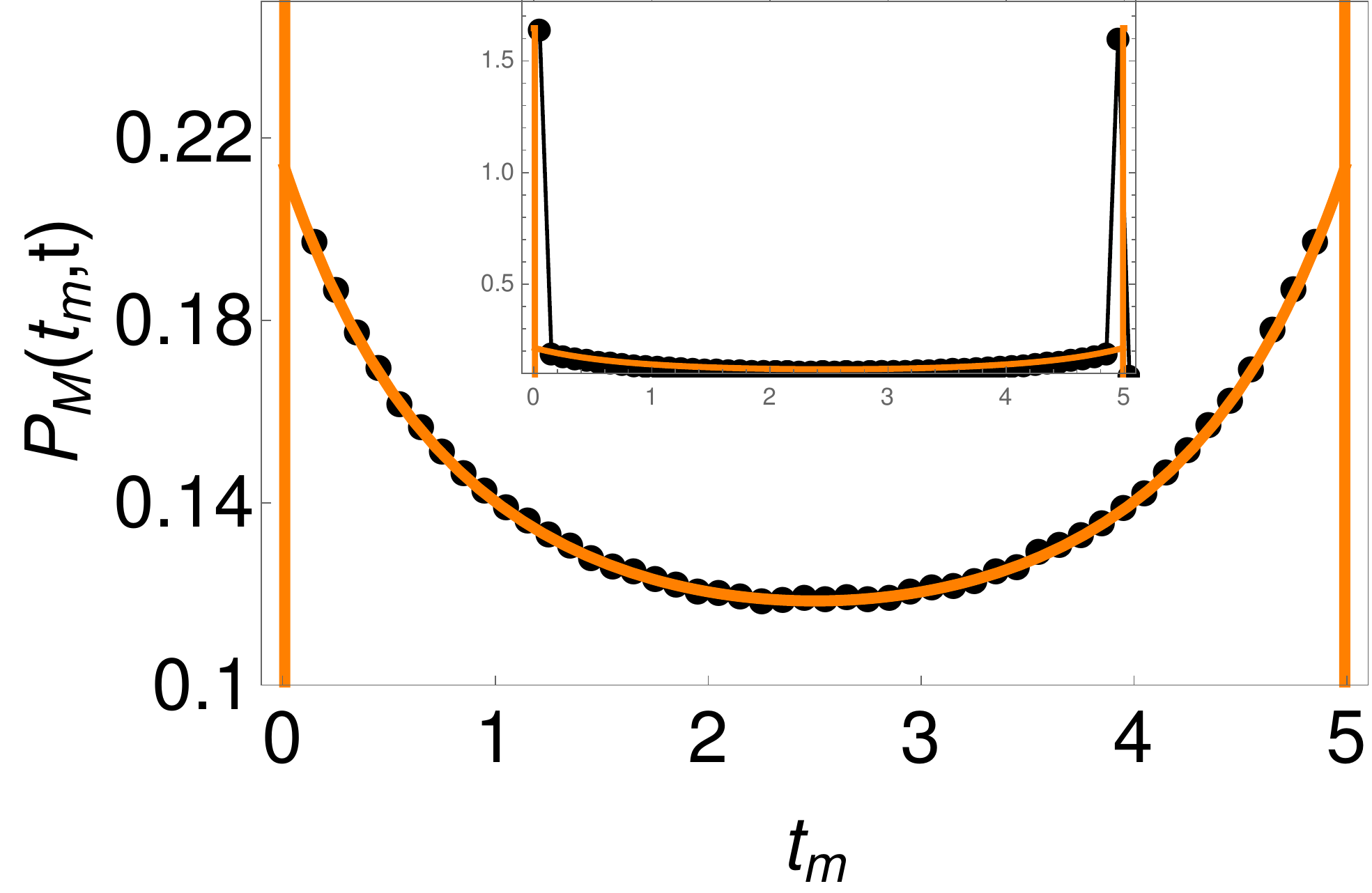}
\centering
\caption{Comparison of analytical expression of $P_M(t_m,t)$ given in Eq.~\eqref{P_t_m} with numerical simulations for $\gamma=1.5$ and $t=5$. The solid orange line is the analytic result and the black dots are obtained in simulations by averaging over $10^6$ realisations. The inset shows the delta functions at $t_m=0$ and $t_m=t$.} 
\label{iprob}
\end{figure}

\subsection{Marginal probability distribution of $t_m$} 
\label{t_m}
We now look at the other marginal distribution $P_M(t_m,t)= \int_0^{vt_m}dM~\mathcal{P}(M,t_m,t)$ of $t_m$. It is easy to perform the integral over $M$ as follows
\begin{align}
P_M(t_m,t)&= \int_0^{vt}dM~\mathcal{P}(M,t_m,t), \nonumber\\
& =\int_0^{vt}dM\Bigg[\frac{h(t)}{2} \delta(M)\delta(t_m) - \frac{\delta(t_m-t) + \gamma~h(t-t_m)}{2} \frac{d}{dM}\left[ \Theta(vt_m-M)~g(M,t_m)\right]\Bigg], \nonumber\\
&= \frac{h(t)}{2} \delta(t_m)+ \frac{\delta(t_m-t) + \gamma~h(t-t_m)}{2} h(t_m), \nonumber\\
&= \frac{h(t)}{2}\Big[\delta(t_m)+\delta(t-t_m) \Big]+\frac{\gamma}{2} h(t-t_m)h(t_m), \label{P_t_m} 
\end{align}
where to go from the second line to the third line, we have used $\int_0^{vt} dM\frac{d}{dM}\left[ \Theta(vt_m-M)\right. $ $\left. g(M,t_m)\right] = -h(t_m) $ with $h(t)$ is given in Eq.~\eqref{S_pm-ep}. 
This is our second main result. Let us try to understand the different terms appearing in the above equation starting with the delta function terms. As mentioned in the previous section,  one can identify that the term proportional to $\delta(t_m)$ gets contributions from trajectories which, starting from the origin have gone to the negative side and never have crossed the origin again within the interval $[0,t]$. On the other hand, the term proportional to $\delta(t_m-t)$ gets contributions from the  trajectories which reach their maxima at the final time $t$ (see fig.~\ref{traj-inversion}a). Interestingly, both these events occur with probability proportional to $h(t)$. For the first term it is easy to understand. For the second term, this can be seen by shifting the origin from $x=0$ to the position of the maxima $x=M$ for each trajectory $\{x(\tau);~0\leq \tau \leq t\}$ before performing the integration over $M$. 
More precisely, if one makes the following variable transformations $\bar{x}=M-x,~\bar{\tau}=t-\tau$ and  $\bar{\sigma}=\sigma$ (as done in \cite{Majumdar10}), then in the new coordinates the RTP,  starting from $\bar{x}=0$ reaches some position above it at  time $t$ while staying above the (new) origin throughout. It is easy to check that the path probabilities of the original process and the transformed process are same. Hence, the survival problem for the original trajectories, starting from some position $x(0)<M$, with the boundary condition $S_+(M,t)=0$ at the absorbing wall placed at $x=M$, is same as the survival problem for the transformed trajectories, starting from $\bar{x}(0)>0$ with the boundary condition $S_-(0,t)=0$ at the absorbing wall placed at $\bar{x}=0$. As a result the survival probability of the transformed trajectories is exactly $h(t)$. 

We now focus on the non-delta function part. This term is proportional to the  product of $h(t_m)$ and $h(t-t_m)$, which are survival probabilities in the first and second durations $t_m$ and $t-t_m$, respectively.  The trajectories that contribute to $t_m$ have their maximum displacements at $t_m$.  Since the process $\{x(\tau),\sigma(\tau)\}$ is Markovian, one can break the whole trajectory into two parts (see fig.~\ref{traj-inversion}b) as done in the previous section. Following \cite{Majumdar10},  we consider the variable transformations $\bar{x}=M-x,~\bar{\tau}=t_m-\tau$ in the first part of duration $t_m$ and $\bar{x}=M-x,~\bar{\tau}=\tau-t_m$ in the second part of duration $(t-t_m)$.  
Under these variable transformations one can see that the contribution from the trajectories in the first and second parts are again survival probabilities $h(t_m)$ and $h(t-t_m)$, respectively.

In Fig.~\ref{iprob} we plot the theoretical result for $P(t_m,t)$ in Eq.~\eqref{P_t_m} along with the same obtained from numerical simulation.  We observe that both the delta function (see inset)  and the non-delta function parts  agree nicely with simulation data. The distribution $P_M(t_m,t)$ for RTP has  interesting feature similar to what one observes for the Brownian particle case. In both cases, the distribution is peaked near $t_m=0$ and $t_m=t$  (see Eq.~\eqref{P_br}). However for RTP, the distribution near the edges do not diverge as in the Brownian particle case,--- instead there are delta functions. It is easy to show that with increasing $\gamma$ the distribution in Eq.~\eqref{P_t_m}  indeed approaches the Brownian motion result in Eq.~\eqref{P_br}.

Note that the distribution $P_{M}(t_m,t)$ in Eq.~\eqref{P_t_m} is computed for the symmetric initial conditions whereby particle starts with $\pm v$ with equal probability. As mentioned earlier, one can however extend this calculation for general initial conditions as shown in \ref{P_M(t_m,t)_assy} where we have provided an explicit expression for $P_{M}(t_m,t)$.

\begin{figure}[t]
\includegraphics[scale=0.4]{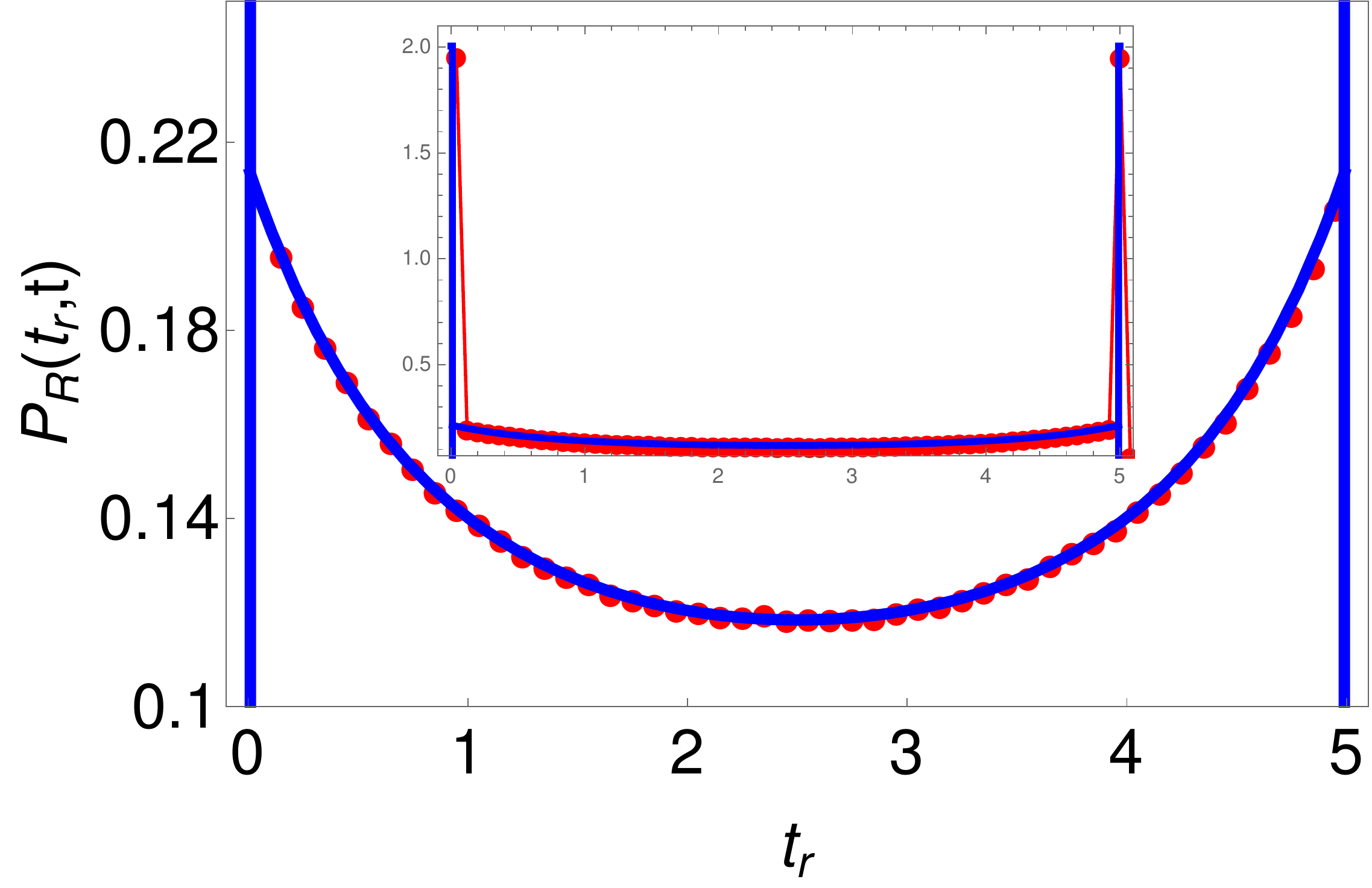}
\centering
\caption{Comparison of analytical expression of $P_R(t_r,t)$ given in Eq.~\eqref{res9} with numerical simulations for $\gamma=1.5$ and $t=5$. The solid blue line is the analytic result and the black dots are obtained in simulations by averaging over $10^6$ realisations. The inset shows the delta functions at $t_r=0$ and $t_r=t$.} 
\label{resfig3}
\end{figure}

\section{The residence time distribution $P_R(t_r,t)$}
\label{disr-t_r}
\noindent
We now study the time that the RTP spends at $x>0$ region over the duration $t$. 
This time, denoted by $t_r$,  is called the residence time and is defined by $t_r=\int^{t}_{0} d\tau \Theta\left(x(\tau)\right)$. The question of the distribution of $t_r$ has been studied for single Brownian particle in one dimension \cite{Majumdar005, Majumdar08, Randon-Furling07}. Intuitively one would naively expect that over a duration $t$, the particle starting at the origin, would spend  half of its time in the $x>0$ region and the remaining half in the $x<0$ region. However, it is known that the probability distribution of $t_r$ (given in Eq.~\eqref{P_br}) is peaked (divergence) near $t_r=0$ and $t_r=t$. This implies that the Brownian particle, once gone to the $+$ve (or $-$ve) side  is reluctant to come back. This behaviour exhibits a counter-intuitive aspect of the Brownian motion. 
In this section, we investigate if an RTP also shows such behaviour in one dimension. 

We want to compute the distribution $P_R(t_r,t)$ of $t_r$ for a given evolution time $t$ when the RTP starts from the origin. To do so we consider the following individual distributions $P^\pm(t_r,t,x_0)$ of $t_r$ corresponding to the initial velocities $\pm v$ of the particle starting at $x_0$, such that $P_R(t_r,t)=[P^+_R(t_r,t,0)+P^-_R(t_r,t,0)]/2$. The factor $1/2$ arises from the probability $1/2$ with which the initial velocity directions are chosen.  Let $Q^{\pm}(p,x_0,t)=\langle e^{-pt_r} |\pm\rangle=\int^{\infty}_{0} dt_r~ e^{-pt_r} P^{\pm}_R(t_r,t,x_0)$ represent the Laplace transformation of $P^\pm_R(t_r,t,x_0)$ with respect to  $t_r$. One can show that $Q^\pm(p,t)$ satisfies the following backward master equations 
\begin{align}
\partial_t Q^{+}(p,x_0,t )&=~~v \partial_{x_0} Q^{+}(p,x_0,t )- \gamma Q^{+}(p,x_0,t )+ \gamma Q^{-}(p,x_0,t )-p U(x_0)Q^{+}(p,x_0,t ),  \label{res6} \\
\partial_t Q^{-}(p,x_0,t )&=-v \partial_{x_0} Q^{-}(p,x_0,t )+ \gamma Q^{+}(p,x_0,t )- \gamma Q^{-}(p,x_0,t )-p U(x_0)Q^{-}(p,x_0,t ),\label{res7} 
\end{align}
where $U(x_0)=\Theta(x_0)$. In \ref{BME-Q} we provide a detailed derivation of the above equations for a general functional $Y[x(\tau)]=\int^{t}_{0} d\tau~U[x(\tau)]$ of the trajectory $x(\tau)$.

We now have to solve the above coupled partial differential equations with appropriate initial and boundary conditions. At $t=0$ the residence time $t_r=0$,  implying $Q^\pm(p,x_0,t=0)=1$. One the other hand at any finite $t$, we have the following boundary conditions:
  \begin{itemize}
      \item If $x_0\to-\infty$, the particle always stays at $x<0$ region over a finite duration $t$ implying $t_r=0$. Hence, $Q^\pm(p,x_0 \to -\infty,t)=1$.
      
      \item If $x_0\to \infty$, the particle is not able to visit the origin in finite time $t$ at $x>0$ implying $t_r=t$. Hence,  $Q^\pm(p,x_0 \to +\infty,t)= e^{-p t}$.  
  \end{itemize}
To proceed further, it seems convenient to take another Laplace transform with respect to $t$ :  $\tilde{ Q}^\pm(p,x_0,s)= \int^{\infty}_0 dt e^{-s t} Q^\pm(p,x_0,t)$. The master equations now become coupled linear first order differential equations, which can be solved with the above mentioned boundary and initial conditions. For better presentation of our results in the main text, here also we prefer to move  the details of the calculation in \ref{P(t_r,t)} while here, present only the expressions of $\tilde{ Q}^{\pm}(p,x_0,s)$ with $x_0=0$:
\begin{align}
\Tilde{Q}^+(p,0,s)=-\frac{1}{2 \gamma}\left(1+\frac{\lambda_0}{s}-\frac{\lambda_p}{s+p}-\frac{\lambda_0 \lambda_p}{s(s+p)} \right), \label{res211} \\
\Tilde{Q}^-(p,0,s)=-\frac{1}{2 \gamma}\left(1-\frac{\lambda_0}{s}+\frac{\lambda_p}{s+p}-\frac{\lambda_0 \lambda_p}{s(s+p)} \right), \label{res222}
\end{align}
where $\lambda_p=\sqrt{(s+p)(2\gamma+s+p)}$ and $\lambda_0=\lambda_{p=0}$. Now to get $P_R(t_r,t)$, one needs to perform the inverse Laplace transform with respect to $s$ and then to $p$ of  $\Tilde{Q}(p,s)=\frac{\Tilde{Q}^+(p,0,s)+\Tilde{Q}^-(p,0,s)}{2}=-\frac{1}{2 \gamma}\left(1-\frac{\lambda_0 \lambda_p}{s(s+p)} \right)$. This can be done by carrying out the appropriate Bromwich integrals. We finally obtain  the following explicit expression for the distribution of $t_r$ spent by a RTP, starting form the origin,  on the positive side over the duration $t$:
\begin{align}
P_R\left(t_r,t \right)&=\frac{ h(t)}{2} \Big[\delta(t_r)+ \delta(t_r-t)\Big]+ \frac{ \gamma}{2}  h(t_r)h(t-t_r),
\label{res9}
\end{align}
where $h(t)$ is given in Eq.~\eqref{S_pm-ep}. 
Note that  this distribution is identical to that of $t_m$ in Eq.~\eqref{P_t_m} as one observes in the context of Brownian motions. The terms involving delta-functions can be interpreted in a similar way as done in the previous section. They get contributions from trajectories which starting from the origin stay either to the $-$ve side or the $+$ve side throughout the entire duration of evolution $t$ and probabilities of such events are $h(t)/2$ each. We observe that, similar to $P_M(t_m,t)$, the non-delta-function part of $P_R(t_r,t)$ is also proportional to the product of $h(t_m)$ and $h(t-t_m)$. Unlike the case of $t_m$, we in this case do not find any obvious connections to survival events associated to $h(t_m)$ and $h(t-t_m)$. The expression in Eq.~\eqref{res9} is verified numerically in the fig.~\ref{resfig3} where we observe very good agreement.

Once again note that the expression for $P_R(t_r, t)$ in Eq.~\eqref{res9} is derived for symmetric initial conditions whereby particle starts with $\pm v$ with equal probability. In this case also one can easily extend the above calculation for general initial conditions as shown in \ref{P(t_r,t)_assy} where we provide an explicit expression for $P_R(t_r, t)$.

\section{Last passage time distribution $P_L(t_\ell,t)$}
\label{dist-t_l}
\noindent
In this section we study the distribution $P_L(t_\ell, t)$ of the time $t_\ell$ at which the RTP crosses the origin for the last time (see Fig.~\ref{fig:traj}) given that it started from the origin choosing $+$ or $-$ direction of the initial velocity with equal probability. Let  $F(a,t)=$ Prob$\left[t_\ell\leq a,t\right]$ be the cumulative distribution of $t_l$ such that $P_L(t_\ell,t)=\left(\partial F_L(a,t)/\partial a \right)_{a-t_\ell}$. It is easy to argue that 
\begin{align}
F_L(a,t)= &2\sum_{\alpha=\pm}\int_0^\infty dy~[\text{prob.~density~to~reach~}y~\text{with~direction~}\alpha ~\text{at~time~}a]~\nonumber \\ 
&~~~~~~~~~~
\times~[\text{Prob.~that~the~particle,~starting~from~}y~\text{with~direction}~\alpha ~\text{does} \nonumber \\ 
&~~~~~~~~~~~~~~~~
\text{never ~come~back~to~the~origin~in~the~remaining~time~}(t-a)] \nonumber \\
=&2\int_0^{va} dy~[P(y,+,a|0,0)~S_+(y,t-a) + P(y,-,a|0,0)~S_-(y,t-a)], \label{F(a,t)}
\end{align}
where $P(y,\pm,a|0,0)=\frac{[P(y,\pm,a|0,+,0)+ P(y,\pm,a|0,-,0)]}{2}$.
Here $P(y,\alpha',t|0,\alpha,0)$ is the probability density with which the RTP, starting from the origin with direction $\alpha \in [+,-]$, reach position $y$ with direction $\alpha' \in [+,-]$ at time $a$. $S_\pm(y,t-a)$ is probability that in the remaining time $(t-a)$, the particle, starting from $y$ with direction $\pm$ did never come back to the origin. This probabilities are basically the survival probabilities as introduced in sec.~\ref{surv-RTP} with the only difference is that, now the absorbing barrier is at the origin instead of at $x=M$. They also satisfy the same backward master equations \eqref{S+d} and \eqref{S-d}. The factor $2$ in Eq.~\eqref{F(a,t)} arises because the RTP can cross the origin (for the last time at $t_\ell<a$) from below or from above and these contributions are equal as can easily be seen by transforming  $y \to -y$ and interchanging $+$ and $-$ directions. Note that, the propagator $P(y,\alpha',t|0,\alpha,0)$ is different from the propagator $P_{\alpha'}(x,t|0,\alpha,0)$ discussed in sec.~\ref{prop-abBC}, although they satisfy same master equations \eqref{P_pm}. The differences are in the boundary conditions. The former one corresponds to the full space problem without any absorbing barrier while the later one corresponds to the case with an absorbing barrier.  The upper limit of the integration in Eq.~\eqref{F(a,t)} is $va$ because that is the maximum distance the RTP can move in time $a$ in the positive direction. 
\begin{figure}[t]
 \includegraphics[scale=0.4]{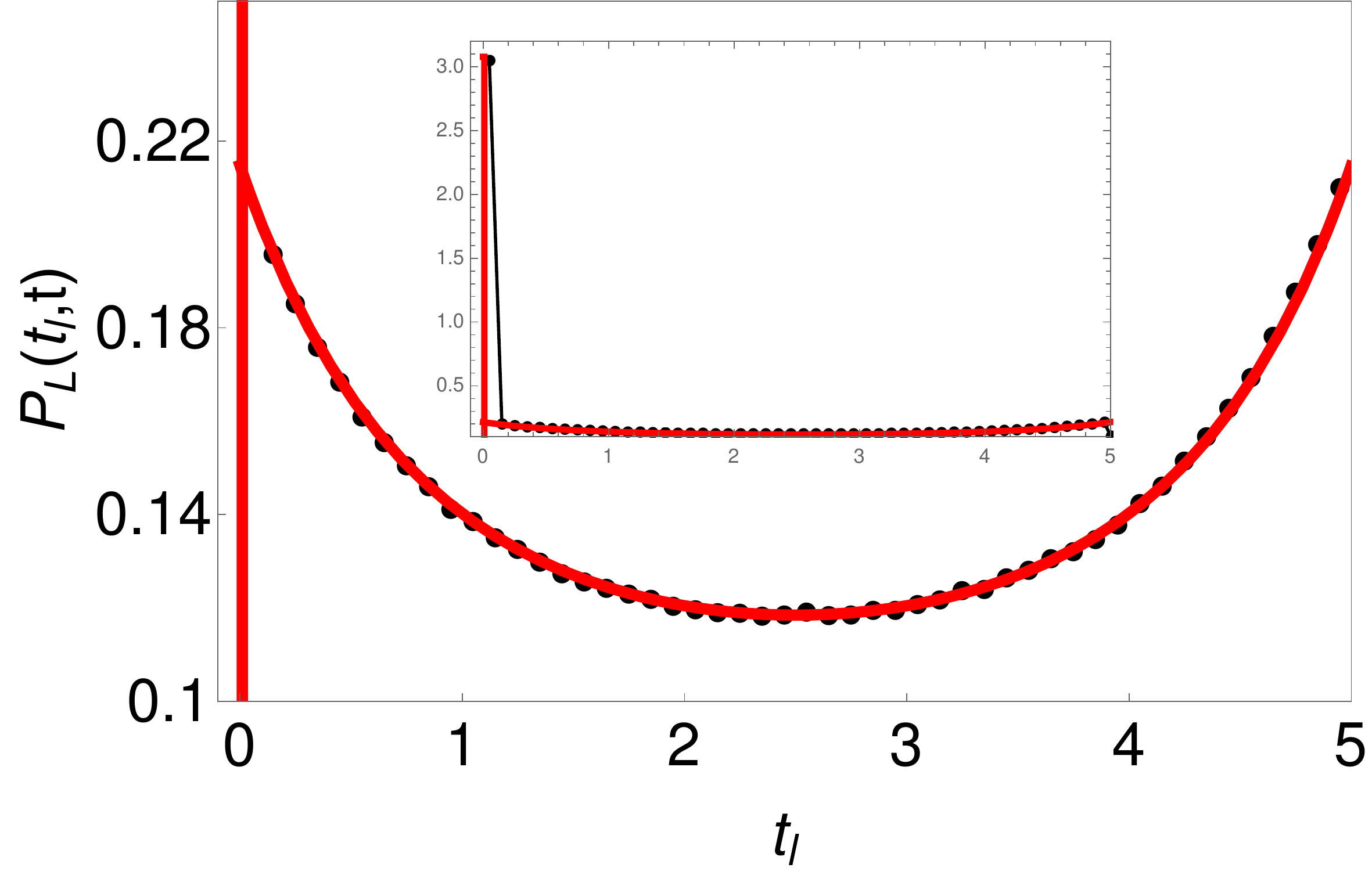}
\centering
\caption{Comparison of analytical expression of $P_L(t_\ell,t)$ given in Eq.~\eqref{lasttime6} with numerical simulations for $\gamma=1.5$ and $t=5$. The solid red line is the analytic result and the black dots are obtained in simulations by averaging over $10^6$ realisations. The inset shows the delta function at $t_\ell=0$.}
\label{lastpic14}
\end{figure}
Now taking derivative with respect to $a$ on both sides of Eq.~\eqref{F(a,t)} and evaluating at $a=t_\ell$ we get
\begin{align}
\begin{split}
P_L(t_\ell,t) ={}& 2v\left[S_+\left(t_\ell,t-t_\ell\right) P(t_\ell,+,t_\ell)+S_-\left(t_\ell,t-t_\ell\right) P(t_\ell,-,t_\ell) \right] \\ 
&+2\int^{vt_\ell}_{0} dy\big[S_+(y,t-t_\ell) \partial _{t_\ell} P(y,+,t_\ell) +S_-(y,t-t_\ell) \partial _{t_\ell} P(y,-,t_\ell) \\
   & -P(y,+,t_\ell)\partial _t S_+(y,t-t_\ell) -P(y,-,t_\ell)\partial _t S_-(y,t-t_\ell)\big],
   \end{split}
\label{lasttime2}      
\end{align}
where we have used short hand notation $P(y,\pm,t)$ for $P(y,\pm,t|0,0)$ by simply $P(y,\pm,t)$. Using Eqs.~\eqref{P_pm}, \eqref{S+d} and \eqref{S-d}, one can simplify the above equation considerably and we finally get
\begin{equation}
 P(t_\ell,t)=2 v S_+(0,t-t_\ell) P(0,+,t_\ell |0,0) , 
\label{lasttime3}
\end{equation}
where we have used the following facts: $P(vt_\ell,-,t_\ell |0,0)=0$ ( because the RTP can not reach the position $vt_\ell$ at time $t_\ell$ with $-$ve velocity) and $S_-(0,t-t_\ell)=0$ (because the absorbing boundary is at the origin). The survival probability can be obtained from Eq.~\eqref{Sm} after putting $x=M$, performing the integral over $\tau$ and finally changing $-$ to $+$. One gets, 
\begin{align}
S_+(0,t-t_\ell)=e^{-\gamma(t-t_\ell)}[I_0(\gamma(t-t_\ell))+I_1(\gamma(t-t_\ell))].
\end{align}
To obtain the $P(0,+,t_\ell)$ we follow the procedure given in \cite{Kanaya18} and get 
\begin{align}
P(0,+,t_\ell) = \frac{\delta(t_\ell)}{2 v} + \frac{\gamma}{4v}[I_0(\gamma t_\ell)+I_1(\gamma t_\ell)].
\end{align}
Inserting, these two expressions in Eq.~\eqref{lasttime2}, we get the following final expression for 
$P(t_\ell,t)$:
\begin{equation}
P_L(t_\ell,t)= \delta(t_\ell) h(\gamma t) + \frac{\gamma}{2} e^{-\gamma t}h(\gamma t_\ell)h(\gamma\left(t-t_\ell\right)),
\label{lasttime6}
\end{equation}
where $h(t)$ is given in Eq.~\eqref{S_pm-ep}. 
Note that the distribution of $P_L(t_\ell,t)$ is almost same as the other two distributions $P_M(t_m,t)$ and $P_R(t_r,t)$ except that the distribution of last passage time (to the origin) has one delta function at $t_\ell=0$. Presence of this $\delta$-function at $t_\ell=0$ stems from the same reasons with which they appear in the cases of $t_m$ and $t_r$. This represents the contributions from those trajectories which starting from the origin, go in either $+$ve or $-$ve direction and stays in that region for the entire duration of evolution $t$.  

One interesting point to note is that although the result in Eq.~\eqref{lasttime3} is derived for the symmetric initial condition case $a=b=1/2$, it turns out that it is valid for arbitrary initial conditions as well. This can be easily verified from the straightforward generalisation of Eq.~\eqref{F(a,t)}.

\section{Conclusion}
\label{conclusion}
In this paper, we mainly have investigated the distributions of three quantities in the context of an RTP moving in one dimension. The three  quantities are, respectively, the time $t_m$ at which maximum displacement $M$ occurs, the time $t_r$ the RTP spends on the $+$ve side and the time $t_\ell$ at which the RTP crosses the origin for the last time. These quantities have been well studied in the context of Brownian motion and other stochastic process, but not well studied for non-Markovian processes except for a few \cite{Sadhu18, Majumdar10, Kasahara77, Majumdar02, Lamperti58} . In the context of random acceleration process \cite{Majumdar10} the distribution of $t_m$ has been computed  however the distribution of other two observables have not been computed. In particular computation of $P_R(t_r,t)$ for random acceleration process is still an open problem.  In the context of Brownian motion, interestingly the distribution of all the three quantities turn out to be identical and their cumulative distributions are given by arcsine function. In this paper we have computed the distributions of $t_m,~t_r$ and $t_\ell$ for an RTP particle starting at the origin $x=0$ with velocities $+v$ (or $-v$) chosen with probability $a$ ( or $b=1-a$). We have shown that these distributions have delta function parts at the edges ($t_c=0$ and $t_c=t$) and a non delta function part. The distributions of $t_m$ and $t_r$ have delta functions at both the edges of the support $[0,t]$ whereas the distribution of  $t_\ell$ has delta function only at one edge $t_\ell=0$. Interestingly, we have found that the non-delta function parts of all the three distributions are  identical for $a=b=1/2$ and are explicitly given by modified Bessel functions. In addition, we also compute the joint distribution $\mathcal{P}(M.t_m,t)$ of $M$ and  $t_m$, which provides us an explicit expression of the marginal distribution of $M$.

The fact that for symmetric initial condition $a=b=1/2$ these distributions are similar (at least the non-delta function parts) is reminiscent of the fact that  in the context of random walks or Brownian motion $B(t)$ the distributions of the three times are identical \cite{feller68}. In the Brownian motion case one can in fact establish that the stochastic times $t_m$ and $t_\ell$ measured over a time duration $t$, are equivalent. This is done based on  the following symmetry properties of Brownian motion: the reflection principle, inversion symmetry [$B(t) \to -B(t)$]  and time reversal symmetry [$\tilde{B}(u)=B(t-u)-B(t)$ is also a Brownian process]. Using these properties one can show that  the stochastic process $X(t)=M(t)-B(t)$ generated by subtracting  $B(t)$ from its current maximum process $M(t)$ is a reflected Brownian motion. For RTP with symmetric initial condition, while the inversion symmetry is present but the reflection symmetry and the time reversal symmetry are not present in the true sense.  As a result it is not obvious if an equivalence between $t_m$ and $t_\ell$ can be established by  extending the arguments of Brownian motion straightforwardly. However, we feel that for RTP the above mentioned symmetry properties are valid approximately which  can possibly explain the similarity between the distributions of $t_m$ and $t_\ell$.



We believe that our results for non-interacting RTPs in one dimension will be informative for addressing the study of $t_m,~t_r$ and $t_\ell$ for other active particle models. For example, often in practical situations the tumbling rate of the bacteria depends on the concentration of the food, which itself can evolve stochastically. Such processes can be described by a RTP whose tumbling rate is coupled to another process. In such cases an interesting statistics to know is the amount of time the bacteria spends in a region of specified food concentration.

\section{acknowledgement}
P. S. and A.K. would like to acknowledge fruitful discussions with Kanaya Malakar, Arghya Das and Urna Basu. 
A.K. would like to acknowledge support from DST under the grant number ECR/2017/000634 and from 
the Indo-French Centre for the promotion of advanced research (IFCPAR) under Project No. 5604-2.

\appendix

\section{The propagator with an absorbing barrier at $x=M$ }
\label{app-propagator}
We start from the coupled differential equations in Eq.~\eqref{P_pm}. We first take Laplace transforms with respect to the time variable:
\begin{equation}
 \bar{P}_\pm(x,s) = \int_0^\infty dt e^{-st}P_\pm(x,t).   
\end{equation}
The differential equations in Eq.(\ref{P_pm}), now becomes 
\begin{align}
\Big(v \partial_x+\gamma +s\Big)\bar{P}_+(x,s)&=~\gamma \bar{P}_-(x,s)+ a\delta(x), \label{Q+} \\
 \Big(v \partial_x-\gamma -s\Big)\bar{P}_-(x,s)&=-\gamma \bar{P}_+(x,s) + b\delta(x), \label{Q-} 
\end{align}   
where the coefficients $a$ and $b$ of the delta functions appear from the initial conditions $P_+(x,0|0,\alpha,0)=a~\delta_{\alpha,+}\delta(x)$ and $P_-(x,0|0,\xi,0)=b~\delta_{\alpha,-}\delta(x)$.
For $x>0$, we can write Eqs.(\ref{Q+}) and (\ref{Q-}) without $\delta(x)$ terms in the RHS,
\begin{align}
L_+\bar{P}_+(x,s)&=\gamma \bar{P}_- (x,s),~\text{with}~~L_+= \Big(v \partial_x+\gamma +s\Big), \label{Q1+} \\
\text{and,}~~~~~ & \nonumber \\
L_-\bar{P}_-(x,s)&=-\gamma \bar{P}_+(x,s), ~\text{with}~~L_-= \Big(v \partial_x-\gamma -s\Big), \label{Q1-} 
\end{align}
Applying $L_-$ on both sides of Eq.~(\ref{Q1+}) and $L_+$ on both sides of Eq.~(\ref{Q1-}), we can write the closed form equations for $\bar{P}_{\pm}$ in the following way:
\begin{align}
\Big(v \partial_x-\gamma -s\Big)\Big(v \partial_x+\gamma +s\Big)\bar{P}_+(x,s)&=-\gamma^2 \bar{P}_+(x,s), \label{Q2+}\\
\Big(v \partial_x+\gamma +s\Big)\Big(v \partial_x-\gamma -s\Big)\bar{P}_-(x,s)&=-\gamma^2 \bar{P}_-(x,s).\label{Q2-}
\end{align}
These two equations are ordinary second-order differential equations and we try the solution $\bar{P}_{+} \sim e^{\beta x}$ and put this in Eq.(\ref{Q2+}). We then get $(\beta-\gamma-s)(\beta+\gamma +s)=-\gamma^2$ which gives $\beta=\pm \frac{1}{v} \sqrt{s(2\gamma +s)}=\pm \lambda(s)$ where $\lambda(s)=\frac{1}{v} \sqrt{s(2\gamma +s)}$. For $x>0^+$, we have $\bar{P}_{+} = A e^{\lambda x}+B e^{-\lambda x}$ where $x$ can go up to $M$. To get $\bar{P}_-$, we can put $\bar{P}_+$ in Eq.(\ref{Q1+}) and get $\bar{P}_-=A(v\lambda +\gamma +s) e^{\lambda x}+ B(-v\lambda +\gamma +s) e^{-\lambda x}$. Similarly, one can also get the solutions of Eqs.(\ref{Q+}) and (\ref{Q-}) for $x<0^-$ region. However, here we will have to consider only positive $\lambda$ so that $\bar{P}_{\pm}$ remain finite as $x  \to -\infty$. We can write the solution of Eqs.(\ref{Q+}) and (\ref{Q-}) as shown below:
\begin{align}
 \bar{P}_+(x,s)&=~~~
    \begin{cases}
      A e^{\lambda(s) x}+ B e^{-\lambda(s) x}, & \text{if}\ 0<x<M \\
      C e^{\lambda(s) x}, & \text{if}\ -\infty<x<0
    \end{cases}
\label{Q++} \\
 \bar{P}_-(x,s)&=\frac{1}{\gamma}
    \begin{cases}
      A(v\lambda(s) +\gamma +s) e^{\lambda(s) x}+ B(-v\lambda(s) +\gamma +s) e^{-\lambda(s) x}, & \text{if}\ 0<x<M \\
      C(v\lambda(s) +\gamma +s) e^{\lambda(s) x}, & \text{if}\ -\infty<x<0
    \end{cases}
\label{Q--} 
\end{align}    
The constants  $A,~B$ and $C$ are still unknown and to determine them we first integrate both sides of 
Eqs.(\ref{Q++}) and (\ref{Q--}) from $x=-\epsilon$ to $x=+\epsilon$ with $\epsilon \to 0^+$. We get the following discontinuity equations
\begin{align} 
  \bar{P}_+(x\to 0^+,s)-  \bar{P}_+(x\to 0^-,s) &=\frac{a}{v},   \label{C1} \\
  \bar{P}_-(x\to 0^+,s)-  \bar{P}_-(x\to 0^-,s) &=-\frac{b}{v}.     
\label{C2} 
\end{align}
We also note that the boundary condition $P_-(M,t)=0$ now reads 
\begin{align}
\bar{P}_-(M,s)=0. \label{Pbar-BC-M}
\end{align}
Inserting the solutions from Eqs.~\eqref{Q++} and \eqref{Q--} in Eqs.~\eqref{C1},~\eqref{C2} and \eqref{Pbar-BC-M}, one gets linear equations for the constants $A,~B$ and $C$ solving which one finally gets these constants in terms of $s,~\gamma,~v$ for different choices of $a$ and $b$.  

\noindent
For example, for $a=1$ and $b=0$ \emph{i.e.} for $+$ initial condition,  one obtains
\begin{align} 
 A(s)&=-\frac{s+\gamma-v\lambda(s)}{2v^2 \lambda(s)} e^{-2M \lambda(s)},~~ \label{C11} \\
 B(s)&= \frac{s+\gamma+v\lambda(s)}{2v^2 \lambda(s)}, ~~~\text{and}, \label{C22} \\
  C(s)&=  \frac{s+\gamma-v\lambda(s)}{2v^2 \lambda(s)} \Big(1-e^{-2M \lambda(s)}\Big).
 \label{C33} 
\end{align}    
Using these expressions of the constants in Eqs.~\eqref{Q++} and \eqref{Q--}  
\begin{align} 
\begin{split}
 \bar{P}_+(x,s|0,+,0)&=\frac{1}{2v^2 \lambda(s)}
    \begin{cases}
    (s+\gamma +v\lambda(s) ) e^{-\lambda(s) x} &- (s+\gamma -v\lambda(s) )  e^{\lambda(s) (x-2M)}, \\
    & ~~~~~~~~~~~~~\text{if}\ 0<x<M \\
      (s+\gamma -v\lambda(s) )&\Big(e^{\lambda(s) x}-  e^{\lambda(s) (x-2M)}\Big), \\ 
      &~~~~~~~~~~~~~ \text{if}\ -\infty<x<0
    \end{cases} 
    \end{split}
\label{Q11} \\
 \bar{P}_-(x,s|0,+,0)&=\frac{\gamma}{2v^2 \lambda(s)}e^{\lambda(s) |x|}-e^{\lambda(s) (x-2M)}.
\label{Q22} 
\end{align}   
Similarly, for $a=0$ and $b=1$   \emph{i.e.} for $-$ initial condition,  one obtains
\begin{align} 
 \bar{P}_+(x,s|0,-,0)&=\frac{\gamma}{2v^2 \lambda(s)} \left(e^{-\lambda(s) |x|}-\frac{s+\gamma-v \lambda(s)}{s+\gamma+v \lambda(s)} \right)e^{\lambda(s) (x-2M)}
\label{Q222} \\
 \bar{P}_-(x,s|0,-,0)&=\frac{1}{2v^2 \lambda(s)}
    \begin{cases}
      (s+\gamma -v\lambda(s) ) e^{-\lambda(s) x}-  (s+\gamma -v\lambda(s) ) &e^{\lambda(s) (x-2M)},\\ 
      & \text{if}\ 0<x<M \\
      (s+\gamma +v\lambda(s) )e^{\lambda(s) x}-(s+\gamma -v\lambda(s) ) &e^{\lambda(s) (x-2M)}, \\
      & \text{if}\ -\infty<x<0.
    \end{cases}
\label{Q333} 
\end{align}    
Now in order to get the solutions in time domain, one need to perform inverse Laplace transforms $\bar{P}_\pm(x,s|0,\pm,0)$.  Looking at the expressions in Eqs.~(\ref{Q11}-\ref{Q333}), we need to know the following inverse Laplace transforms 
\begin{align} 
&L^{-1}_{s\to t}\left[ \frac{1}{\lambda(s)} e^{-g\lambda(s)} \right]= ve^{-\gamma t} I_0\left(\frac{\gamma}{v}\sqrt{v^2t^2-g^2}\right) 
\Theta(vt-g) \label{I} \\
\begin{split}
&L^{-1}_{s\to t} \left [ \frac{s+\gamma +\alpha~v \lambda(s)}{\lambda(s)} e^{-\lambda(s) g} \right]
= \gamma v e^{-\gamma t} \sqrt{\frac{vt-g}{vt+g}}I_1\left(\frac{\gamma}{v}\sqrt{v^2t^2-g^2}\right)\Theta(vt-g)  \\
&~~~~~~~~~~~~~~~~~~~~~~~~~~~~~
- \frac{1+\alpha}{2} v^2 \frac{d}{d g}\left[ \Theta(vt-g) I_0\left(\frac{\gamma}{v}\sqrt{v^2t^2-g^2}\right) \right],~~
\text{for}~~\alpha=\pm1, 
\end{split}
\label{j} \\
\begin{split}
&L^{-1}_{s\to t}\left[ (s+\gamma-v \lambda(s)) e^{-\lambda(s) g} \right]= 
\frac{\gamma v~e^{-\gamma t}}{vt-g} \left[\frac{\gamma g}{v} I_0\left(\frac{\gamma}{v}\sqrt{v^2t^2-g^2}\right) \right. \\ 
&~~~~~~~~~~~~~~~~~~~~~~~~~~~~~~~~~~~~~~~~~~~~~~~~
\left.+\sqrt{\frac{vt-g}{vt+g}}I_1\left(\frac{\gamma}{v}\sqrt{v^2t^2-g^2}\right)\right]\Theta(vt-g).
\end{split}
 \label{T}
\end{align}    
Using these results and performing some simplifications, one arrives at the expressions of the propagators given in Eqs.~(\ref{Pm_p}-\ref{Pp_m})

\section{Cumulative distribution $Q(M,t)$}
\label{cum_dist}
The cumulative distribution for $M$ is defined as,
\begin{align}
Q(M,t)=\int_0^MdM'~\mathcal{P}(M',t)=\int_0^MdM' \int_0^tdt_m~\mathcal{P}(M',t_m,t),
\label{cum1}
\end{align}
where $\mathcal{P}(M,t_m,t)$ is given by Eq.~\eqref{P_joint}. One can easily integrate Eq.~\eqref{P_joint} with $t_m$ to get $\mathcal{P}(M,t)$. 
\begin{align}
\begin{split}
\mathcal{P}(M,t)&=\int_0^tdt_m~\mathcal{P}(M',t_m,t)  \\
&=\frac{h(t)}{2} \delta(M)-\frac{1}{2}\frac{d}{dM}\left[ \Theta(vt-M)~g(M,t)\right] \\
&~~~-\frac{\gamma}{2}
\int_0^t dt_m h(t-t_m) \frac{d}{dM}\left[ \Theta(vt_m-M)~g(M,t_m)\right].
\end{split}
\label{cum2}
\end{align}
The contribution of first term in the R.H.S of Eq.~\eqref{cum2} to Eq.~\eqref{cum1} is just $\frac{h(t)}{2}$. The contribution from the second term is given by,
\begin{align}
\frac{1}{2}\int_0^M dM'\frac{d}{dM'}&\left[ \Theta(vt-M')~g(M',t)\right]= \frac{1}{2}\left[ \Theta(vt-M')~g(M',t)\right]\bigg|^M_0 ,\nonumber \\
&=\Theta\left(vt-M\right)\frac{g\left(M,t\right)}{2}-\frac{h(t)}{2}.
\label{cum3}
\end{align} 
Similarly one can get the contribution of the third term in the R.H.S of Eq.~\eqref{cum2} to Eq.~\eqref{cum1}.
\begin{align}
\frac{\gamma}{2}\int_0^M dM'\int_0^t dt_m h(t-t_m) \frac{d}{dM}&\left[ \Theta(vt_m-M)~g(M,t_m)\right] \nonumber \\ 
&=\frac{\gamma}{2}\int_0^t dt_m h(t-t_m)\left[ \Theta(vt_m-M')~g(M',t_m)\right]\bigg|_0^M, \nonumber\\
&=\frac{\gamma}{2}\int_0^t dt_m h(t-t_m)\left[\Theta\left(vt_m-M\right)g\left(M,t_m\right)-h(t_m)\right],\nonumber\\
&=\frac{\gamma}{2}\int_0^t dt_m \Theta\left(vt_m-M\right)h(t-t_m)g\left(M,t_m\right) -\left[1-h(t) \right],\nonumber
\label{cum4}
\end{align}
where we have used Eq.~\eqref{inte1} to go from second line to third line. Adding all these contributions, we finally get $Q(M,t_m)$,
\begin{align}
Q(M,t)=1-\frac{1}{2}g(M,t)\Theta(vt-M) -\frac{\gamma}{2}\int_0^tdt'~\Theta(vt'-M)~h(t-t')~g(M,t').
\end{align}

\section{Proof of the identity}
\label{integration}
\noindent
Here we prove the identity 
\begin{align}
\gamma \int_0^t dt' h(t-t') h(t') \stackrel{?}{=}2[1-h(t)].
\label{inte1}
\end{align}
where $h(t)=e^{-\gamma t}[I_0(\gamma t)+I_1(\gamma t)]$.
The best way to prove this  identity is to take the Laplace transform of the left hand side with respect to 
the variable $t$. Since the left hand side is in the convolution form, we have 
\begin{align}
L_{t\to s}\left[\gamma \int_0^t dt' h(t-t') h(t')\right]=\gamma \tilde{h}(s)^2.
\label{inte3}
\end{align}
where $\tilde{h}(s)$ is the Laplace transform of $h(t)$ and is given by
\begin{align}
\tilde{h}(s)=L_{t\to s}[h(t)]=\frac{1}{\gamma}\left(\frac{\sqrt{s(2\gamma+s)}}{s}-1\right)..
\label{inte2}
\end{align}
Using this result in Eq.~\eqref{inte3} and performing some algebraic simplifications, we get 
\begin{align}
L_{t\to s}\left[\gamma \int_0^t dt' h(t-t') h(t')\right] =2\left[\frac{1}{s}-\frac{1}{\gamma}\left(\frac{\sqrt{s(2\gamma+s)}}{s}-1\right) \right].
\label{inte4}
\end{align}
Finally performing inverse Laplace transform on both sides with respect to $s$ one proves the identity in Eq.~\eqref{inte1}.


\section{Backward master equation for $Q^\pm(p,x_0,t)$}
 \label{BME-Q}
Here we provide a derivation of the Backward master equations in \eqref{res6} and \eqref{res7} for a general functional $Y\left[ x\left( t\right)\right]$ 
\begin{align}
 Y\left[ x\left( t\right)\right]= \int^t_0 d\tau U\left[ x\left( \tau\right)\right], \label{res1}
\end{align}
of the trajectory $\{x(\tau);~0 \leq \tau \leq t \}$ of  the RTP starting from $x_0 $ at $\tau =0$. Here we mainly follow the general steps done for Brownian motion in \cite{Majumdar005}. It is clear that $Y$ is a random variable. Let $P^{\sigma}(Y,x_0,t)$ denote the probability density for $Y$ where $\sigma=\pm1$ is the direction of the initial velocity. We define the characteristic function:
\begin{align}
  Q^{\sigma}(p,x_0,t)=\int^{\infty}_0 dY e^{-p Y} P^{\sigma}\left(Y,x_0,t\right) =\left< e^{-p Y}\right>_{\left(x_0,\sigma\right)},
\label{res2}
\end{align}
 where $<...>_{\left(x_0,\sigma\right)}$ denotes the average with initial position $x_0$ and velocity orientation $\sigma$. Following the definition in Eq.(\ref{res2}), we write for a small time interval $\Delta t$
\begin{align}
\begin{split}
  Q^{\sigma}\left(p,x_0,t+\Delta t \right)=&\left< e^{-p \int^{t+\Delta t}_0 d\tau U\left[ x\left( \tau\right)\right]}\right>_{\left(x_0,\sigma\right)}\\
  &=\left< e^{-p \int^{\Delta t}_0 d\tau U\left[ x\left( \tau\right)\right]}\indent e^{-p \int^{t+\Delta t}_{\Delta t} d\tau U\left[ x\left( \tau\right)\right]} \right>_{\left(x_0,\sigma\right)}\\
  &=\left(1-p U(x_0)\Delta t\right)\left<e^{-p \int^{t+\Delta t}_{\Delta t} d\tau U\left[ x\left( \tau\right)\right]} \right>_{\left(x_0,\sigma\right)}.
\label{res3}
\end{split}
\end{align}
where $(x_0, \sigma)$ is the initial configuration.  In time interval $\Delta t$ the state $(x_0, \sigma)$ of the RTP can change  to $(x_0+\sigma v \Delta t, \sigma)$ with probability $1-\gamma \Delta t$ and to $(x_0, -\sigma)$ with probability $\gamma \Delta t$. We can write Eq.(\ref{res3}) as following:

\begin{align}
  Q^{\sigma}(p,x_0,t+\Delta t )& =\left(1-p U(x_0)\Delta t\right)\left[\left(1-\gamma \Delta t\right)\langle e^{-p \int^{t+\Delta t}_{\Delta t} d\tau U\left[ x\left( \tau\right)\right]} \rangle_{\left(x_0+\sigma v \Delta t, \sigma\right)} \right. \\
&\left. +\gamma \Delta t\left<e^{-p \int^{t+\Delta t}_{\Delta t} d\tau U\left[ x\left( \tau\right)\right]} \right>_{\left(x_0-\sigma v \Delta t, -\sigma\right)}\right]\\  &=\left(1-p U(x_0)\Delta t\right)\left[ \left(1-\gamma \Delta t\right) Q^{\sigma}(p,x_0+\sigma v \Delta t,t )+\gamma \Delta t Q^{-\sigma}(p,x_0,t )\right].
\end{align}
Taking the $\Delta t \to 0$ limit we obtain 
\begin{align}
\partial_t Q^{\sigma}(p,x_0,t )=\sigma v \partial_{x_0} Q^{\sigma}(p,x_0,t )- \gamma Q^{\sigma}(p,x_0,t )+ \gamma Q^{-\sigma}(p,x_0,t )-p U(x_0)Q^{\sigma}(p,x_0,t ).
\label{res5}
\end{align}
If we choose $U(x)= \Theta(x)$ then $Y$ represents the residence time $t_r$ the particle on the positive semi axis  and for this choice one obtains the differential equations in Eq.(\ref{res6}) and Eq.(\ref{res7}).

\section{Probability distribution $P_R(t_r,t)$ }
\label{P(t_r,t)}
In this appendix, we find the solutions of Eqs.(\ref{res6}) and (\ref{res7}) to obtain $P_R(t_r,t)$. We take the Laplace transformation of $Q^{\pm}(p,x_0, t)$ in the time $t$ as $\tilde{ Q}^{\pm}(p,x_0,s)= \int^{\infty}_0 dt e^{-s t} Q^{\pm}(p,x_0,t)$. As a result Eqs.(\ref{res6}) and (\ref{res7})now become 
\begin{align}
\left[v \partial_{x_0}-\gamma-pU(x_0)-s\right]\Tilde{Q}^+\left(p,x_0,s\right) &=-\gamma \Tilde{Q}^-\left(p,x_0,s\right)-1,
\label{res11}\\
\left[v \partial_{x_0}+\gamma+pU(x_0)+s\right]\Tilde{Q}^-\left(p,x_0,s\right) &=\gamma \Tilde{Q}^+\left(p,x_0,s\right)+1.
\label{res12}
\end{align}
The boundary conditions for $Q^{\pm}\left(p,x_0,t\right)$ [discussed in section(\ref{disr-t_r})] can now be translated for $\Tilde{Q}^{\pm}\left(p,x_0,s\right)$ as $\Tilde{Q}^{\pm}(p,x_0\to-\infty,s)=\frac{1}{s}$ and $\Tilde{Q}^{\pm}(p,x_0\to\infty,s)=\frac{1}{s+p}$. Let us first solve the Eqs.(\ref{res11}) and (\ref{res12}) for $x_0>0$ region where we put $U(x_0)=1$ and we have,
\begin{align}
\left[v \partial_{x_0}-\gamma-p-s\right]\Tilde{Q}^+\left(p,x_0,s\right) &=-\gamma \Tilde{Q}^-\left(p,x_0,s\right)-1,
\label{res21}\\
\left[v \partial_{x_0}+\gamma+p+s\right]\Tilde{Q}^-\left(p,x_0,s\right) &=\gamma \Tilde{Q}^+\left(p,x_0,s\right)+1.
\label{res22}
\end{align}
These equations can be made homogenous by shift $Q^{\pm}\left(p,x_0,s\right)=\frac{1}{s+p}+Z^{\pm}\left(p,x_0,s\right)$ to get,
\begin{align}
\left[v \partial_{x_0}-\gamma-p-s\right]Z^+\left(p,x_0,s\right) &=-\gamma Z^-\left(p,x_0,s\right),
\label{res13}\\
\left[v \partial_{x_0}+\gamma+p+s\right]Z^-\left(p,x_0,s\right) &=\gamma Z^+\left(p,x_0,s\right).
\label{res14}
\end{align}
The boundary condition of $Q^{\pm}(p,x_0 \to \infty,s)=\frac{1}{s+p}$ now becomes $Z^{\pm}(p,x_0 \to -\infty,s)=0$. Following the same procedure as done in sec.~\ref{app-propagator}, we find that the solutions are given by
\begin{align}
\Tilde{Q}^+_+\left(p,x_0,s\right)&=\frac{1}{s+p}+ A(s,p) e^{-\frac{\lambda_p(s) x_0}{v}},
\label{res17}\\
&~~~~~~~~~~~~~~~~~~~~~~~~~~~~~~~~~~~~~~~~~~~~~~~~~~~~~~~~~~~~~~~~
\text{for},~~~x_0>0, \nonumber \\
\Tilde{Q}^-_+\left(p,x_0,s\right)&=\frac{1}{s+p}+ \frac{A(s,p)}{\gamma} \left(s+p+\gamma +\lambda_p \right) e^{-\frac{\lambda_p(s) x_0}{v}},
\label{res18}
\end{align}
where $\lambda_p(s)=\sqrt{\left(s+p\right)\left(2\gamma+s+p\right)}$. 
The solutions obtained in Eqs.(\ref{res17}) and (\ref{res18}) are true for $x_0>0$ region. 
For $x_0<0$, one can follow the same procedure. Only difference is that now $U(x_0)=0$. We find 
\begin{align}
\Tilde{Q}^+\left(p,x_0,s\right)&=\frac{1}{s}+ B(s,p) e^{\frac{\lambda_0(s) x_0}{v}},
\label{res15}\\
&~~~~~~~~~~~~~~~~~~~~~~~~~~~~~~~~~~~~~~~~~~~~~~~~~~~~~~~~~~~~~~~~
\text{for},~~~x_0<0, \nonumber \\
\Tilde{Q}^-\left(p,x_0,s\right)&=\frac{1}{s}+ \frac{B(s,p)}{\gamma} \left(s+\gamma -\lambda_0(s) \right) e^{\frac{\lambda_0(s) x_0}{v}}.
\label{res16}
\end{align}
 The constants $A(s,p)$ and $B(s,p)$ in the expressions for $\Tilde{Q}^\pm_{\pm}\left(p,x_0,s\right)$ are unknown constants which can be computed from the continuity of $\Tilde{Q}^\pm\left(p,x_0,s\right)$ at $x=0$, \emph{i.e.} $\Tilde{Q}^{\pm}(x_0 \to 0_+)=\Tilde{Q}^{\pm}(x_0 \to 0_-)$. After some simple manipulations, we find
\begin{align}
A(s,p)&=\frac{p\left(\lambda_0(s)-s \right)}{s(s+p)(p+\lambda_0(s)+\lambda_p(s))}.
\label{res20} \\
B(s,p)&=-\frac{p\left(s+p+\lambda_p(s) \right)}{s(s+p)(p+\lambda_0(s)+\lambda_p(s))},
\label{res19}
\end{align}
For $x_0=0$ the solutions simplifies further and we have 
\begin{align}
\Tilde{Q}^+\left(p,0,s\right)=-\frac{1}{2 \gamma}\left(1+\frac{\lambda_0(s)}{s}-\frac{\lambda_p(s)}{s+p}-\frac{\lambda_0 \lambda_p(s)}{s(s+p)} \right),
\label{res21} \\
\Tilde{Q}^-\left(p,0,s\right)=-\frac{1}{2 \gamma}\left(1-\frac{\lambda_0(s)}{s}+\frac{\lambda_p(s)}{s+p}-\frac{\lambda_0(s) \lambda_p(s)}{s(s+p)} \right).
\label{res22}
\end{align}
If the particle starts with $\pm v$ velocity with equal probability $1/2$, the total probability is given by $P_R(t_r,t)=\frac{P^+_R(t_r,t)+P^-_R(t_r,t)}{2}$. Accordingly, $\tilde{ Q}(p,0,s)= \int^{\infty}_0 dt e^{-s t} \int_0^t dt_r e^{-pt_r}P_R(t_r,t)= \frac{\Tilde{Q}^+(p,0,s)+\Tilde{Q}^-(p,0,s)}{2}$ which, after some simplifications, is given by
\begin{align}
\Tilde{Q}(p,0,s)&=\frac{\Tilde{Q}^+(p,s)+\Tilde{Q}^-(s,p)}{2}=-\frac{1}{2 \gamma}\left(1-\frac{\lambda_0(s) \lambda_p(s)}{s(s+p)} \right).
\label{res23}
\end{align}
Now to get $P_R(t_r,t)$ we have to perform inverse Laplace transform of  $\Tilde{Q}(p,s)$ with respect to $s$ and $p$ both \emph{i.e.}. 
\begin{align}
P_R(t_r,t)=L^{-1}_{s\to t} L^{-1}_{p\to t_r} \left[-\frac{1}{2 \gamma}\left(1-\frac{\lambda_0(s) \lambda_p(s)}{s(s+p)} \right) \right].
\label{res24}
\end{align}
Using the following inverse Laplace transform result,
\begin{align}
L^{-1}_{p\to t_r} \left[\frac{\lambda_p(s)}{s+p} \right]= \delta(t_r)+\gamma e^{-s t_r} e^{-\gamma t_r} \left[I_0(\gamma t_r)+I_1(\gamma t_r) \right]
\label{res25}
\end{align}
we get the final solution as given in Eq.~(\ref{res9}) of the main text.


\section{Probability distribution for time $t_m$ to reach maximum $M$ for general initial condition }
\label{P_M(t_m,t)_assy}

In section (\ref{t_m}) we computed the distribution for the time to reach maximum $t_m$ for a RTP that starts from $x=0$ with $\pm v$ with equal probability \emph{i.e.} $a=b=1/2$. Here we derive the distribution  $P_M^{as}(t_m, t)$ of $t_m$ for arbitrary $a$ and $b$. ($'as'$ in the superscript is to represent assymetric initial conditions). We follow the same procedure as done in sec.~\ref{prop-abBC}. Using the expressions for $P_{\pm}(M-\epsilon,t_m|0,\pm,0)$ obtained in section (\ref{prop-abBC}), we first compute  $p_{\pm}(M-\epsilon,t_m|0,0)=aP_{\pm}(M-\epsilon,t_m|0,+,0)+bP_{\pm}(M-\epsilon,t_m|0,-,0)$ for general initial conditions. 
After some simplifications, the expressions for $p_{\pm}(M-\epsilon,t_m|0,0)$ read as,
\begin{align}
&\left.\begin{aligned} 
p_+(M-\epsilon,t_m|0,0)  &\approx  -\frac{d}{dM}\left[ \Theta(vt_m-M)~g^{as}(M,t_m)\right] + O\left(\epsilon \right),     \\
p_-(M-\epsilon,t_m|0,0)   &\approx  -\frac{\gamma~\epsilon}{v} \frac{d}{dM}\left[ \Theta(vt_m-M)g^{as}(M,t_m)\right]+O(\epsilon^2),  
       \end{aligned}
 \right\}
 \label{P_pm_ep_assy}
 \end{align}
where the function $g(M,t)$ in Eq.~\eqref{g(M,t)} is now changed to
\begin{align}
 g^{as}(M,t)=e^{-\gamma t}\left[aI_0\left(\frac{\gamma}{v}\sqrt{v^2t^2 -M^2}\right)+b\sqrt{\frac{v t-M}{ vt+M}} I_1\left(\frac{\gamma}{v}\sqrt{v^2t^2 -M^2}\right) \right].
 \label{g(M,t)_assy}
\end{align}
Using these expressions in Eq.~\eqref{JPdf-nd}, we get the part of the joint distribution $\mathcal{P}^{as}(M,t_m,t)$ of $M$ and $t_m$ for $0<M<vt$ and $0<t_m<t$. 

\begin{figure}[t]
\includegraphics[scale=0.25]{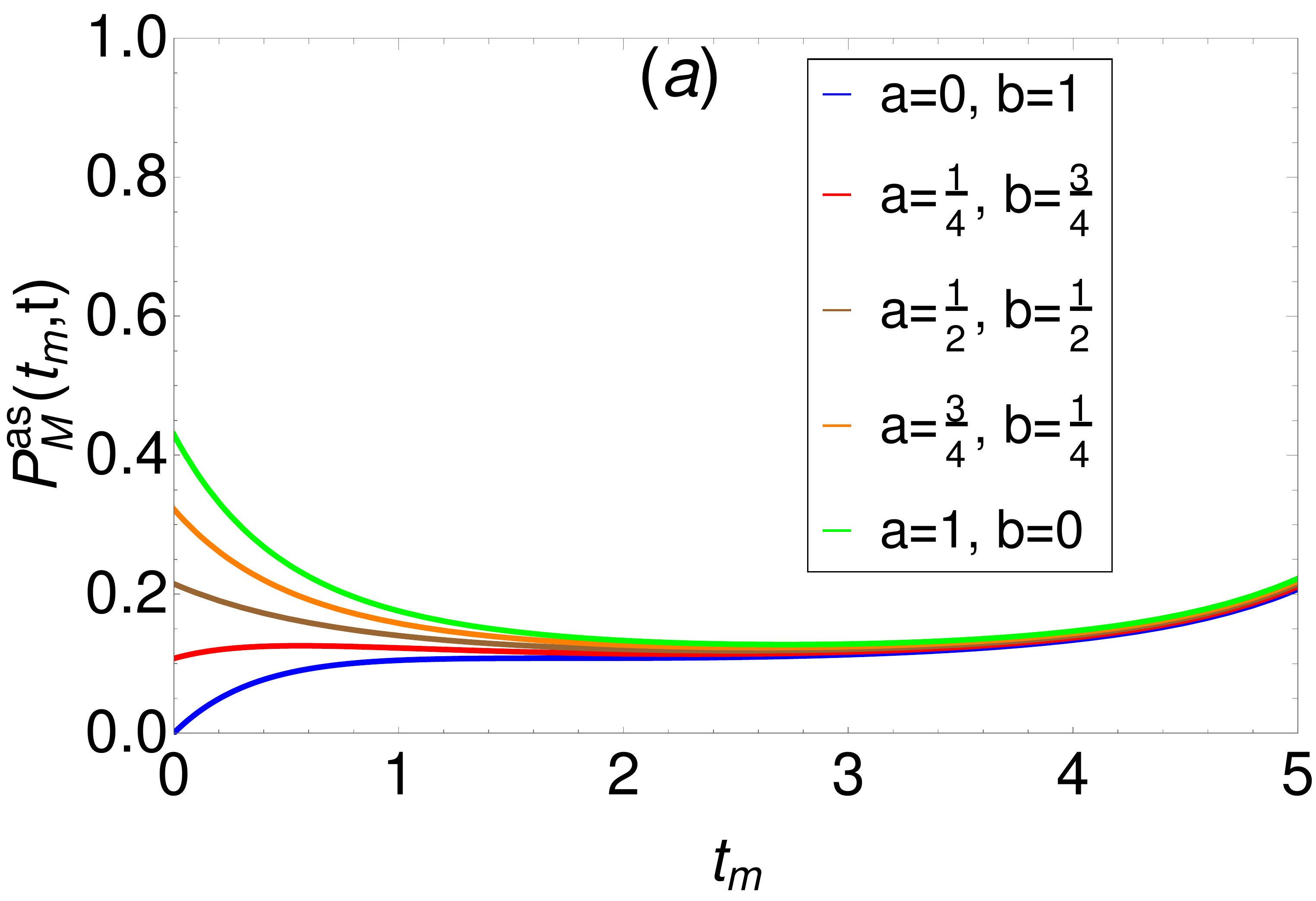}
\includegraphics[scale=0.3]{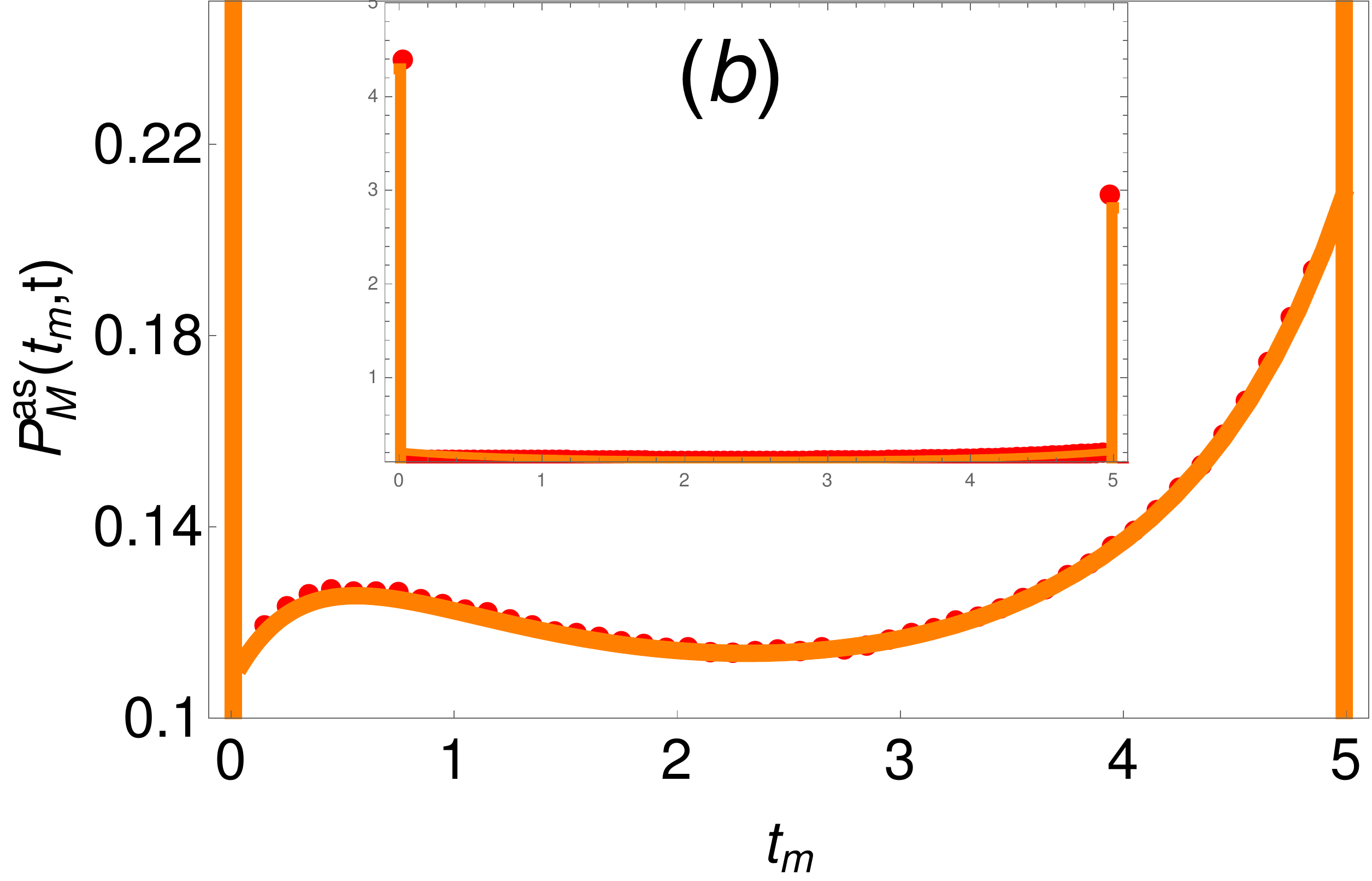}
\centering
\caption{(a) Plot of the $P_M^{as}(t_m,t)$ in Eq.~\eqref{P_t_m_assy} for various combinations of $(a,b)$. Note that we have not shown delta-function part here. (b) Numerical verification of the same for $a = 0.25, b=0.75$. The inset shows the comparision of the delta-function parts. The Other parameters used in both the plots are $\gamma=1.5$ and  $t=5$.}
\label{max_assy_pic}
\end{figure}

We now have to add the contribution from the events $M=0$ (or equivalently $t_m=0$) and $t_m=t$. 
As discussed in the symmetric case, this contribution from the events $t_m=0$ comes from those paths which initially moves with $-v$ and survives the origin till time $t$. Hence, 
\begin{align}
\underbrace{\mathcal{P}^{as}(M,t_m,t)}_{\text{contribution~from~paths~with~}t_m=0} &= \delta(t_m) \delta(M) b h(t).
\end{align}
On the other hand the contributions to $t_m=t$ event come from trajectories which starting from the origin reach the maximum displacement $M$ at time $t_m = t$.  
The contribution  is obtained by generalising Eq.~\eqref{pam} straightforwardly for general initial conditions as 
\begin{align}
\underbrace{\mathcal{P}^{as}(M,t_m,t)}_{\text{contribution~from~paths~with~}t_m=t} = \delta(t_m-t)~[a P_+(M,t|0,+,0)+ b P_+(M,t|0,-,0)].
\label{pam_assy}
\end{align}
Inserting the explicit forms of these propagator $P_+(M,t|0,\pm,0)$ from Eqs.~\eqref{Pp_p} and \eqref{Pp_m}, and simplifying we get 
\begin{align}
\underbrace{\mathcal{P}^{as}(M,t_m,t)}_{\text{contribution~from~paths~with~}t_m=t} 
&=- \frac{d}{dM}\left[ \Theta(vt-M)~g^{as}(M,t)\right],
\end{align}
Now putting all the three contributions together, performing some simplifications, fixing the normalisation and integrating over $M$, we finally get,
\begin{align}
P_M^{as}(t_m,t)&= \int_0^{vt}dM~\mathcal{P}^{as}(M,t_m,t), \nonumber\\
&= b h(t)\delta(t_m)+h^{as}(t)\delta(t-t_m) +\gamma h(t-t_m)h^{as}(t_m), \label{P_t_m_assy} 
\end{align}
where $h^{as}(t)=e^{-\gamma t}[a I_0(\gamma t)+b I_1(\gamma t)]$. Note that for $a=b=\frac{1}{2}$, $h^{as}(t)=\frac{h(t)}{2}$ and Eq.~\eqref{P_t_m_assy} reduces to the symmetric case distribution in Eq.~\eqref{P_t_m}. However for $(a \neq b)$ ,$P_M^{as}(t_m,t)$ is different from Eq.~\eqref{P_t_m}. In Fig.~\ref{max_assy_pic}a, we have plotted our analytical result for various combinations of $a$ and $b$ and in Fig.~\ref{max_assy_pic}b we have verified it with numerical simulations for a particular choice of $a$. We observe  excellent agreement between the two.

\section{Probability distribution for residence time $t_r$ for general initial condition}
\label{P(t_r,t)_assy}

\begin{figure}[t]
 \includegraphics[scale=0.3]{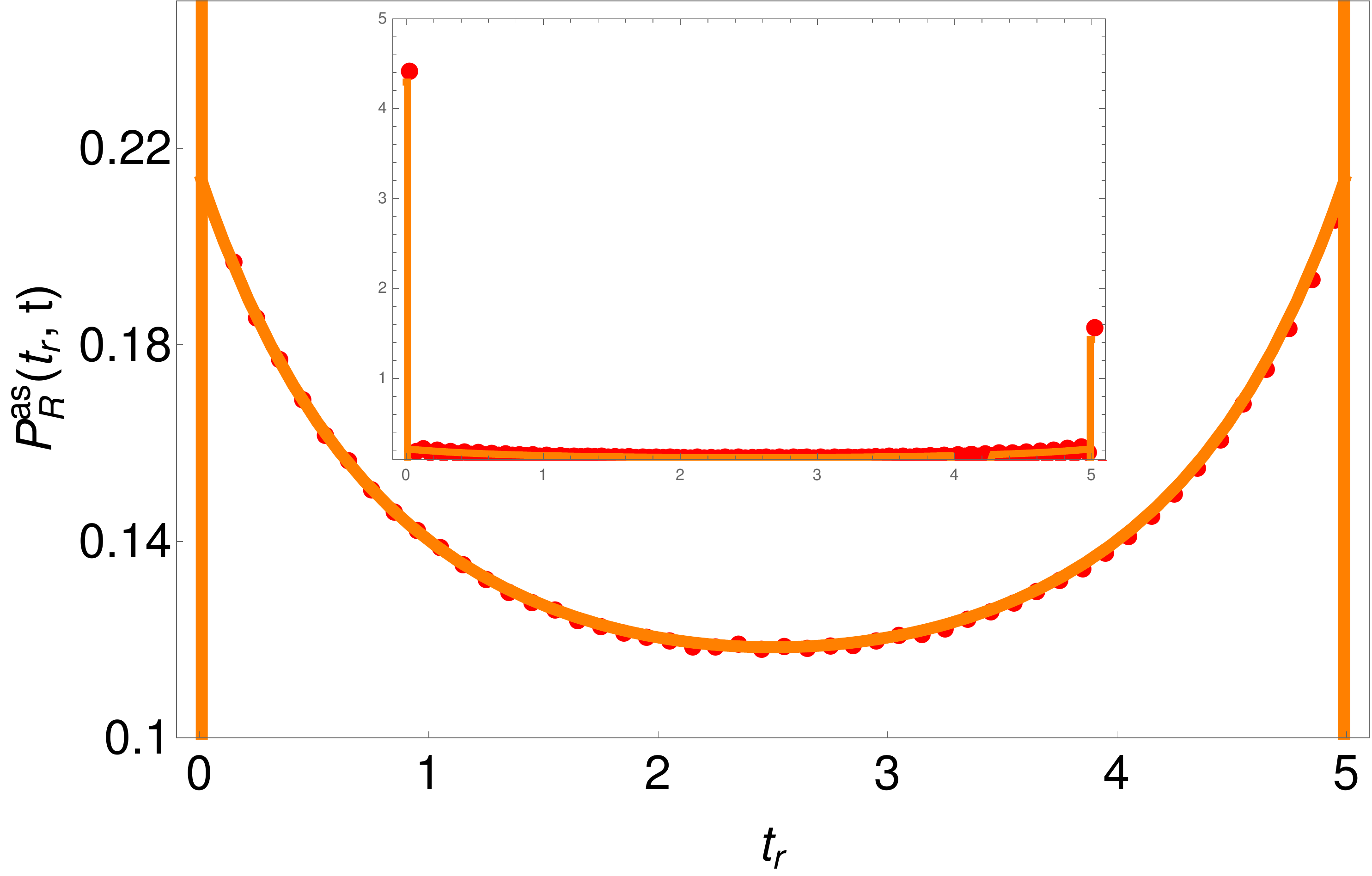}
\centering
\caption{Comparison of analytical expression of $P_R^{as}(t_r,t)$ given in Eq.~\eqref{residence_2} with numerical simulations for $a = 0.25, b=0.75, \gamma=1.5$ and $t=5$. The solid red line is the analytic result and the black dots are obtained in simulations by averaging over $10^7$ realisations. The inset shows the delta function at $t_r=0$ and $t_r=t$. Note that the non-delta function part is identical to that in Fig(\ref{resfig3}) for symmetric case but the delta-function parts are different.}
\label{assy_residence_pic1}
\end{figure}

In this section, we extend the computation of the residence time distribution $P_R(t_r,t)$ in section (\ref{disr-t_r}) from symmetric initial condition to asymmetric initial condition for the velocity.  For this we assume that the particle starts with $+ v(-v)$ velocity with probability $a (b)$ such that $a+b=1$. Let us denote the Laplace transform of $P_R^{as}(t_r, t)$ with respect to $t$ and $t_r$ by $\Tilde{Q}^{as}(p,s)$ where  $'as'$ in the superscript once again represents asymmetric initial condition. Using the Eqs.~\eqref{res211} and \eqref{res222}, and performing some simplifications we get 
\begin{align}
\Tilde{Q}^{as}(p,s)&= a \Tilde{Q}_{+}(p,s)+b \Tilde{Q}_{-}(p,s), \nonumber\\
& =-\frac{1}{2 \gamma}\left(1-\frac{\lambda_0 \lambda_p}{s(s+p)} \right)-\frac{a-b}{2\gamma} \left(\frac{\lambda_0}{s}-\frac{\lambda_p}{s+p} \right). \label{residence_1} 
\end{align}
To get the distribution $P_R^a(t_r, t)$ in the time domain, we have to do two Laplace transformations: one from $s \to t$ and the other from $p \to t_r$. Looking at the expression in Eq.~\eqref{residence_1}, one notes that the first term in R.H.S is identical to the symmetric case whose Laplace transformation is given in Eq.~\eqref{res9}. The Laplace transformation of the second term in R.H.S of Eq.~\eqref{residence_1} is given in  Eq.~\eqref{res25}. One can therefore write the distribution for $t_r$ for general initial condition as
\begin{align}
P_{R}^{as}(t_r, t)&=  h(t) \Big[b \delta(t_r)+ a \delta(t_r-t)\Big]+ \frac{ \gamma}{2}  h(t_r)h(t-t_r).\label{residence_2} 
\end{align}
Note that $P_{R}^{a}(t_r, t)$ in Eq.~\eqref{residence_2} reduces to $P_{R}(t_r, t)$  in Eq.~\eqref{res9} for $a=b=\frac{1}{2}$. Also note that the asymmetry in the initial condition only alters the strengths of the delta-functions of $P_{R}(t_r, t)$ but leaves the non-delta function part unchanged. These weights of the delta-functions can be understood in the same way as for the symmetric case. In Fig.(\ref{assy_residence_pic1}), we verify our analytical result with numerical simulation and find excellent agreement with it.  

\end{document}